\documentclass[a4paper,10pt,twocolumn,preprint,3p]{elsarticle}
\usepackage{color}
\usepackage{xcolor}
\usepackage[english]{babel}
\usepackage[utf8]{inputenc}

\usepackage{rakmacros}
\usepackage{multirow}

\title{Simulating vacuum arc initiation by coupling emission, heating and plasma processes}
\date{\today}

\usepackage{orcidlink}

\begin{document}

\begin{frontmatter}

\journal{Vacuum}

\author[a,b] {Roni Koitermaa\orcidlink{0000-0001-9814-7358}}%[orcid=0000-0001-9814-7358]
\author[a,b] {Andreas Kyritsakis\orcidlink{0000-0002-4334-5450}\corref{cor1}}%[orcid=0000-0002-4334-5450]
\ead{andreas.kyritsakis@ut.ee; akyritsos1@gmail.com}
\cortext[cor1]{Corresponding author}
\author[b] {Tauno Tiirats\orcidlink{0000-0002-1256-7392}}%[orcid=0000-0002-1256-7392]
\author[b] {Veronika Zadin\orcidlink{0000-0003-0590-2583}}%[orcid=0000-0003-0590-2583]
\author[a] {Flyura Djurabekova\orcidlink{0000-0002-5828-200X}}%[orcid=0000-0002-5828-200X]

\affiliation[a] {
  organization={Helsinki Institute of Physics and Department of Physics, University of Helsinki},
  addressline={P.O. Box 43 (Pietari Kalmin Katu 2)},
  postcode={FI-00014},
  city={Helsinki},
  country={Finland}
}
\affiliation[b] {
  organization={Institute of Technology, University of Tartu},
  addressline={Nooruse 1},
  postcode={50411},
  city={Tartu},
  country={Estonia}
}

\renewcommand\floatpagefraction{.7}

\selectlanguage{english}

\begin{abstract}
Vacuum arcing poses significant challenges for high-field vacuum devices, underscoring the importance of understanding it for their efficient design. A detailed description of the physical mechanisms involved in vacuum arcing is yet to be achieved, despite extensive research. In this work, we further develop the modelling of the physical processes involved in the initiation of vacuum arcing, starting from field emission and leading to plasma onset. Our model concurrently combines particle-in-cell with Monte Carlo collisions (PIC-MCC) simulations of plasma processes with finite element-based calculations of electron emission and the associated thermal effects. Including the processes of evaporation, impact ionization and direct field ionization allowed us to observe the dynamics of plasma buildup from an initially cold cathode surface. We simulated a static nanotip at various local fields to study the thresholds for thermal runaway and plasma initiation, identifying the significance of various interactions. We found that direct field ionization of neutrals has a significant effect at high fields on the order of 10 GV/m. Furthermore, we find that cathode surface interactions such as evaporation, sputtering and bombardment heating play a major role in the initiation of vacuum arcs. Consequently, the inclusion of these interactions in vacuum arc simulations is imperative.
\end{abstract}


\end{frontmatter}

\section{Introduction}\label{sec:introduction}

Vacuum arcing, also known as vacuum breakdown (VBD), has been of interest for many decades both for its engineering applications and for the intricate physics involved. High-field vacuum devices such as particle accelerators, vacuum interrupters and X-ray sources experience limitations caused by the uncontrolled appearance of vacuum arcing \cite{clic, boxarc}. Similarly, frequent vacuum arcs in accelerating structures limit the performance of the CERN Compact Linear Collider (CLIC), designed to achieve unprecedentedly high radio-frequency electromagnetic field gradients to accelerate electron and positron beams \cite{clic}.

Despite many experimental and computational studies, a full understanding of the phenomenon is yet to be achieved. Formation of vacuum arcs involves the coupling of many physical processes that act on different time and length scales. This imposes certain challenges for detailed studies of this complex phenomenon. Computational simulations, on the other hand, are able to couple different processes within a single model, providing powerful means for studies on the mechanisms of plasma initiation and its subsequent evolution. For instance, the recent computational effort \cite{arcpic} has considered the dynamics of plasma onset by coupling surface dynamics and plasma physics, while the thermal runaway process was explained by coupling electron and atomic dynamics with the electrostatic field, updated on-the-fly following the field-emitting tip evolution \cite{femocs}.

It is widely accepted that intense field emission of electrons leading to local heating and metal atom evaporation is the main precursor process that leads to a vacuum arc. Emission of electrons is determined both by the temperature of the tip and the surface electric field, so emission is coupled to both the heating of the cathode and the space charge distribution around the cathode. Metal atoms evaporate from the cathode as a result of heating. These atoms can ionize to form plasma, which can conduct current between the cathode and the anode \cite{zhoexp}. The positive plasma ions can in turn impact the cathode surface and sputter neutral atoms which feed the plasma in a positive feedback loop until the process reaches a self-sustained steady state \cite{arcpic}.

In this work, we focus on the initial stages of vacuum arc onset near a field-emitting tip by developing an accurate physical model for plasma-surface interactions, which leads to the formation of a fully self-sustained plasma near the tip surface. In our model, a nanotip or nanoprotrusion is assumed to be already present on the cathode surface. The aspect ratio of the assumed tip is sufficient for the electric field enhancement to trigger electron emission. Although the nature of the appearance of such protrusions is not fully understood, the proposed mechanisms to explain formation of field-emitting tips link them to biased surface diffusion under electric field gradients \cite{jandif, kimdif} as well as to subsurface plastic deformations around near-surface defects \cite{pohdlo, zaddlo}. Furthermore, it has been proposed that breakdown crater edges could serve as precursor sites for subsequent breakdowns \cite{sarexp}.
Regardless of the origin of field-enhancing protrusions, understanding the processes that lead from field emission to plasma initiation is of paramount importance for mitigating vacuum arc occurrence. These processes determine the dynamic coupling of the developing arc to the electromagnetic power available in the system, which is empirically found to determine vacuum arc occurrence  \cite{wuepow, paspow, grupow, wanpow}. Hence, an accurate description of these processes can facilitate the optimization of high-field devices for arc mitigation.

Previously, simulations of vacuum arc initiation have been performed using the ArcPIC \cite{arcpic} and FEMOCS codes \cite{femocs}. ArcPIC focused on plasma simulation, but with a simplified description of plasma-surface interactions. The flux of neutral atoms was linked to the electron currents and no dynamic heating was taken into account. FEMOCS was used to model emission and heating effects, but without full plasma simulation. Hence, to achieve understanding of the dynamic interplay of the processes contributing to plasma buildup, it is important to combine both approaches in a single model. In this work, we add plasma simulation to the FEMOCS code, and simulate the thermal runaway process leading to plasma buildup in a static nanotip. These simulations aim to describe more accurately the plasma-surface interactions that contribute towards the formation of a vacuum arc, as well as to study the impact of different physical interactions on the process of plasma buildup.

\section{Methods}\label{sec:methods}

\newcommand{\cu}{\x{Cu}}

\newcommand{\pa}{\x{A}} % particle A
\newcommand{\pb}{\x{B}} % particle B
\newcommand{\qa}{q_\pa} % charge
\newcommand{\qb}{q_\pb}
\newcommand{\za}{z_\pa} % charge
\newcommand{\zb}{z_\pb}

\newcommand{\ma}{m_\pa} % mass
\newcommand{\mb}{m_\pb}

\newcommand{\ud}{\mathcal{U}} % uniform distribution
\newcommand{\nd}{\mathcal{N}} % normal distribution

\subsection{FEM solution of electric field and heating}

\fig{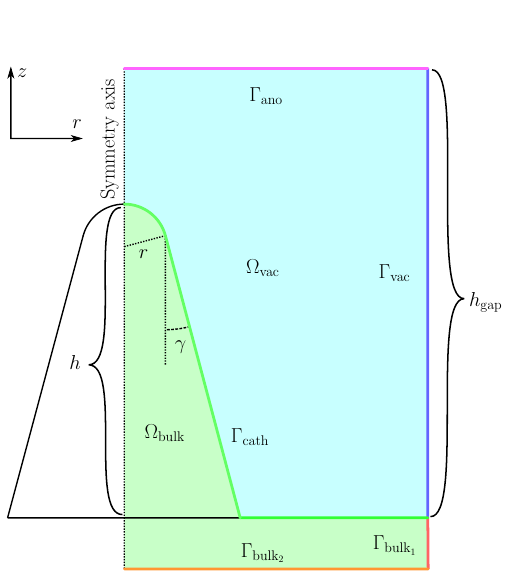}{Simulation geometry (vacuum $\Omega_\x{vac}$ and bulk $\Omega_\x{bulk}$) and boundaries $\Gamma$.\label{fig:domains}}{0.8}

We model the arc gap as a 2D axisymmetric domain as shown in figure \ref{fig:domains}. On the cathode surface (green) $\Gamma_\x{cath}$, we place a nanotip around which the local electric field is enhanced. At the anode (magenta) $\Gamma_\x{ano}$, we apply a DC voltage $V_0$, while the cathode surface $\Gamma_\x{cath}$ is electrically connected to the bulk bottom (orange) $\Gamma_{\x{bulk}_2}$, which is grounded. This results in an electric field in the vacuum (blue) $\Omega_\x{vac}$. The electric field as well as the heating of the cathode are found using the finite element method (FEM).

When a voltage is applied between the anode and the cathode, an electric field forms in the vacuum between them. We assume the electrostatic case, where the electric field is not changing with time and the electron currents are relatively small such that the effects of magnetic fields can be neglected. Moreover, it has been observed in experiments that vacuum breakdowns in radio frequency (RF) electromagnetic fields exhibit similar properties as those that appear in the direct current (DC) electrostatic case. This supports the hypothesis that the underlying physical mechanisms are the same at the microscopic level \cite{desarc}.

We find the electric field in the vacuum above the cathode by solving \textit{Poisson's equation}
\be
\nabla^2 \phi = -\frac{\rho}{\eps_0},\label{eq:poi}
\ee
where $\phi$ is the electric potential, $\rho$ is the charge density and $\eps_0$ is the vacuum permittivity. The boundary conditions in our system are
\begin{subequations}
\begin{align}
\phi(\vb r) = V_0, \ &\vb r \in \Gamma_\x{ano}, \label{eq:bcf1}\\
\nabla \phi(\vb r) \cdot \vb n = 0, \ &\vb r \in \Gamma_\x{vac}, \label{eq:bcf2}\\
\phi(\vb r) = 0, \ &\vb r \in \Gamma_\x{cath},\label{eq:bcf3}
\end{align}
\end{subequations}
where $V_0$ is the voltage at the anode $\Gamma_\x{ano}$ and $\vb n$ is the cathode surface normal. The electric field on the vacuum boundary $\Gamma_\x{vac}$ and the potential on the cathode surface $\Gamma_\x{cath}$ are zero.

Thermal field emission of electrons from the surface of the cathode causes current in the bulk $\Omega_\x{bulk}$, which must be conserved at every point. The current distribution can be found from the \textit{continuity equation}
\be
\nabla \cdot (\sigma(T) \nabla \phi_J) = 0, \label{eq:cont}
\ee
where $\sigma(T)$ is the electric conductivity and $\phi_J$ is the (current) potential. The boundary conditions are
\begin{subequations}
\begin{align}
\sigma(T) \nabla \phi_J(\vb r) = -\vb J_e, \ &\vb r \in \Gamma_\x{cath}, \label{eq:bce1}\\
\sigma(T) \nabla \phi_J(\vb r) = 0, \ &\vb r \in \Gamma_{\x{bulk}_1}, \label{eq:bce2}\\
\phi_J(\vb r) = 0, \ &\vb r \in \Gamma_{\x{bulk}_2},\label{eq:bce3}
\end{align}
\end{subequations}
i.e. the potential on the bulk bottom boundary $\Gamma_{\x{bulk}_2}$ and the current density on the bulk right boundary $\Gamma_{\x{bulk}_1}$ are zero, while the current density $\vb J_e$ on the cathode surface $\Gamma_\x{cath}$ is given by electron emission.

Currents inside the nanotip generate resistive heating, increasing the temperature of the nanotip. The resulting temperature $T$ can be found by solving the \textit{heat equation}
\be \label{eq:heat}
C_v \frac{\partial T}{\partial t} = \nabla \cdot (\kappa(T) \nabla T) + \frac{J^2}{\sigma(T)},
\ee
where $J$ is the current density in the bulk, $C_v$ is the volumetric heat capacity and $\kappa(T)$ is the thermal conductivity. The temperature-dependent thermal conductivity is obtained from the Wiedemann-Franz law \cite{wiefra}
\be
\kappa(T) = L T \sigma(T),
\ee
where we use the Lorentz number $L = 2.0 \tp{-8} \us{W} \ \Omega \usp{K}{-2}$ \cite{natlon}. The heat equation has the boundary conditions \cite{femocs}
\begin{subequations}
\begin{align}
\kappa(T) \nabla T(\vb r) \cdot \vb n = P_\x{cath}, \ &\vb r \in \Gamma_\x{cath}, \label{eq:bct1}\\
\kappa(T) \nabla T(\vb r) \cdot \vb n = 0, \ &\vb r \in \Gamma_{\x{bulk}_1}, \label{eq:bct2}\\
T(\vb r) = T_\x{amb}, \ &\vb r \in \Gamma_{\x{bulk}_2},\label{eq:bct3}
\end{align}
\end{subequations}
where $T_\x{amb}$ is the ambient temperature. The heat flux on the cathode surface is given by the collective contribution from the Nottingham effect, cooling due to evaporation and heating by plasma ion bombardment:
$P_\x{cath} = P_N + P_\x{vap} + P_\x{bom}$. For the cathode $\Omega_\x{bulk}$, we have the initial condition $T(t=0) = T_\x{amb}$.

The 2D axisymmetric geometry of the simulation domain has rotational symmetry along the $z$-axis of the nanotip. The bulk and vacuum domains are discretized into quadrilaterals forming a mesh with a density that increases towards the top of the nanotip. The greatest density of particles will be seen at the top of the nanotip, so we need the greatest mesh density in this region of interest. The weak forms of equations \ref{eq:poi}, \ref{eq:cont} and \ref{eq:heat} are derived in \cite{femocs}, which are discretized using the Galerkin method to get a system of equations that can be solved numerically. Discretization of the simulation domains is done using GMSH \cite{gmsh}, while the discretized equations are solved using deal.II \cite{dealii} in FEMOCS.

\subsection{PIC-MCC particle simulation}

To simulate the initiation of plasma around the nanotip, we use the \textit{particle-in-cell} (PIC) method with \textit{Monte Carlo collisions} (MCC). The particle-in-cell method is used to calculate macroscopic quantities on a stationary mesh resulting from the distribution of particles \cite{birpic, pic}, e.g. to determine the charge density distribution. In the particle-in-cell method, multiple particles can be represented as \textit{superparticles} (SPs), which have the same charge-to-mass ratio as single particles. This way, a much larger number of particles can be simulated. The particles are simulated in a 2D axisymmetric vacuum mesh. The left boundary has this axisymmetric boundary condition, so particles are reflected if they cross it (figure \ref{fig:domains}). If particles cross the boundaries $\Gamma_\x{cath}$, $\Gamma_\x{vac}$ or $\Gamma_\x{ano}$, they are removed from the simulation. The 2D3V method is used to add a third velocity component for particles \cite{apman}. At each step, particles moving in the $\phi$ direction are rotated back into the $rz$ plane. At each time step, we advance the positions of particles using the leapfrog integration scheme. Details of this implementation are available in \cite{femocs}.

To model interactions between particles, we perform
collisions between them. Each cell in the mesh contains
a number of particles, and collisions can occur randomly
between particles in each cell. The process of how particles are randomly matched for a collision is described in \cite{takcol}. For different particle species, we randomly collide ``projectile'' SPs (with lower number density in a cell) with ``target'' SPs (with larger number density in a cell). The collision probability $P \in [0, 1)$ is estimated from the collision cross section according to \cite{mccpic}
\be \label{eq:colp}
P = 1 - \exp \lb -u n \sigma(E) \Delta t \rb,
\ee
where $u = |\mathbf{u}_\pa - \mathbf{u}_\pb|$ is the magnitude of relative velocity between particles A and B, $n = \max(n_A, n_B)$ is the SP number density of the target particle species inside a cell, $\sigma(E)$ is the energy-dependent cross section of a specific collision and $\Delta t$ is the time step. The collision is considered to occur when a uniformly distributed random number satisfies $U \sim \ud(0, 1) < P$.

New velocities after the collision are calculated by finding a velocity difference in a random direction. We assume that the collisions are isotropic. For this, we use the rotation matrix $R$ \cite{takcol}
\be
R(\theta, \phi) = \bpm
\cos \theta \cos \phi & \cos \theta \sin \phi & -\sin \theta \\
-\sin \phi & \cos \phi & 0 \\
\sin \theta \cos \phi & \sin \theta \sin \phi & \cos \theta
\epm
\ee
with scattering angle $\theta$ and azimuth angle $\phi$. These are randomly selected as $\phi' \sim \ud(0, 2 \pi)$ and $\cos \theta' \sim \ud(-1, 1)$.
The velocity difference is
\be \label{eq:du}
\Delta \vb u = R^{-1}(\theta, \phi) \lb R^{-1}(\theta', \phi') - I \rb (0, 0, u)^\T.
\ee
The post-collision velocities of the colliding particles are $\vb u' = \vb u + \Delta \vb u$.

Elastic collisions between particles A and B (notation below is used for collisions)
\be
\pa + \pb \rightarrow \pa + \pb
\ee
can be described using the binary collision approximation. Probability of this collision is calculated using equation \ref{eq:colp}.
The resulting velocity change is applied to the two colliding particles as \cite{matcol}
\begin{subequations}
  \label{eq:newv}
  \begin{align}
    \vb v_\pa' = \vb v_\pa + \mu / \ma \Delta \vb u, \\
    \vb v_\pb' = \vb v_\pb - \mu / \mb \Delta \vb u,
  \end{align}
\end{subequations}
where $\mu = \frac{\ma \mb}{\ma + \mb}$ is the effective mass, $\vb v_\pa$ is the original velocity of the colliding particle A and $\vb v_\pa'$ is the velocity after collision. Monte Carlo collisions for other types of collisions such as ionization and charge exchange can be performed following a similar approach \cite{matcol}. Impact ionization produces an ion from the impact of an electron and neutral: $\cu + \el \rightarrow \cu^+ + 2 \el$. A summary of collision processes is given in table \ref{tab:colls}. In collisions such as ionization, we split SPs when necessary, which is described in section \ref{section:dynwei}. The details of our implementation are described in \cite{koimsc}.

\subsection{Coupling plasma simulation to emission and heating calculations}

\subsubsection{Simulation model}

\fig{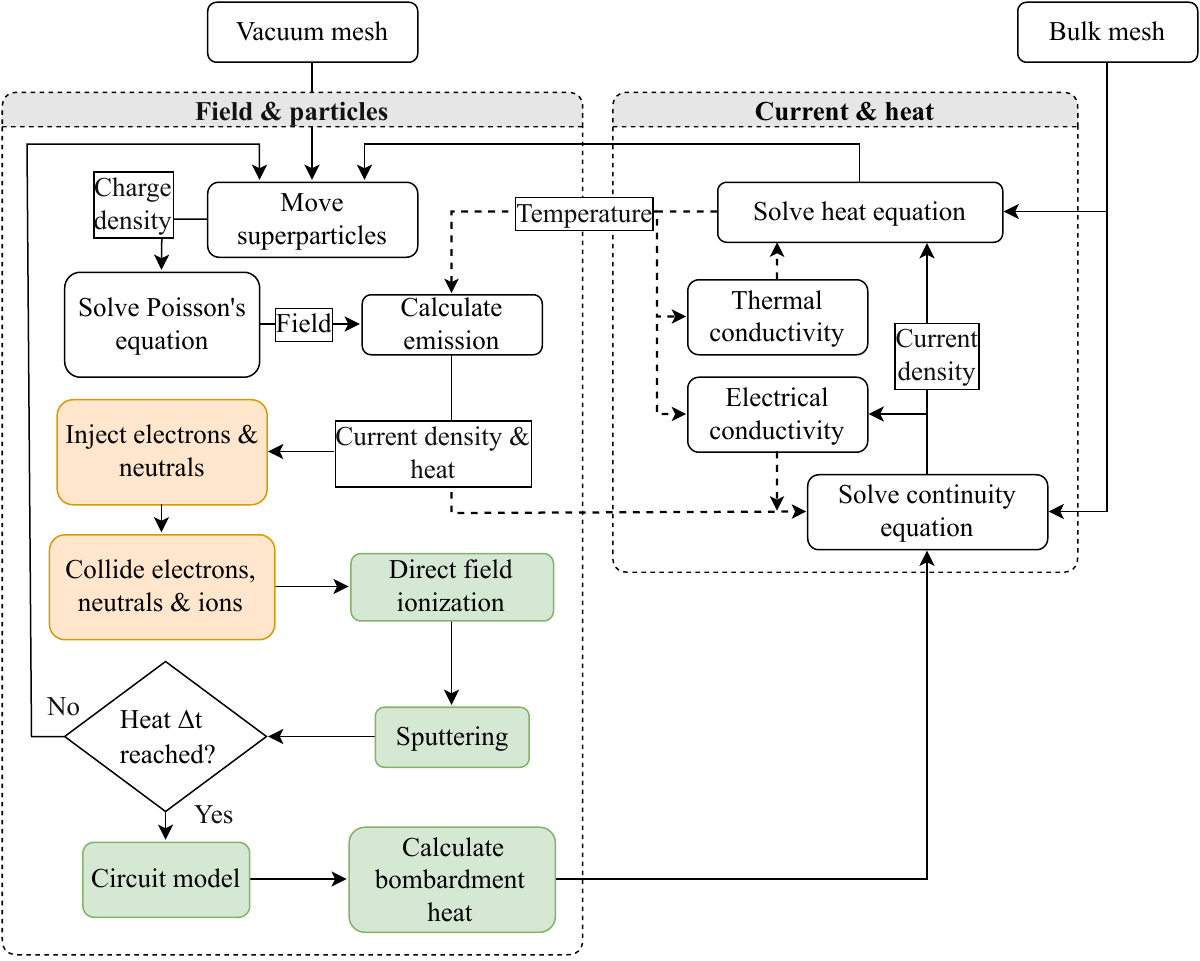}{Simulation model with changes highlighted.\label{fig:newpic}}{1.0}
The development of plasma around the nanotip is interlinked with the the dynamics of the runaway process. To understand which of the factors promote the runaway process and which may potentially suppress it, we must couple the simulations of plasma development directly to simulations of electron emission and current-generated heating processes. Each of these aspects of vacuum arcing has been studied separately in previous works, namely using the ArcPIC code to study plasma development \cite{arcpic} and the FEMOCS code to study thermal runaway \cite{femocs}. In this work, we introduce the PIC methodology for plasma simulation into FEMOCS, as well as extending the model to include other processes such as field ionization, sputtering and bombardment heating to better capture the dynamics of surface interactions. A schematic of our current simulation model is shown in figure \ref{fig:newpic}, where added steps are shown in green and modified steps in orange.

We calculate electron emission currents from the cathode surface $\Gamma_\x{cath}$ in the form of the general thermal field model \cite{jengtf, eimgtf}, which includes both regimes of field and thermionic emission. For calculation of electron emission current, as well as for calculation of the corresponding Nottingham heat $P_N$, we use the GETELEC code \cite{getelec}. This code automatically calculates the electron emission in each of the emission regimes: the cold field emission, thermionic emission and intermediate regime. Based on the emission current at a given location on the cathode surface, we determine the number of electron SPs that are injected at this location into the plasma calculations.

The number of neutrals that evaporate from the cathode surface is calculated using as in our previous work \cite{zhoeva} using the \textit{Clausius-Clapeyron equation}
\begin{equation}
    \phi_\x{vap} = \frac{p_\x{atm}}{\sqrt{2 \pi m k_B T}}  \exp \lsb \frac{E_\x{vap}}{k_B} \lb \frac{1}{T_b} - \frac{1}{T} \rb \rsb, 
\end{equation}
where $\phi_\x{vap}$ is the flux of evaporated neutrals, $p_\x{atm}$ is the atmospheric pressure, $E_\x{vap}$ is the energy required to evaporate an atom (3.43 eV for Cu), $T_b$ is the boiling point of Cu and $k_B$ is the Boltzmann constant. Neutrals are injected with speeds sampled from the Maxwell-Boltzmann distribution based on the surface temperature.
Additionally, the evaporation of neutrals leads to evaporative cooling of the cathode surface. Considering that each evaporation event absorbs on average $E_\x{vap} + k_B T$ (potential and kinetic energy), the heat flux is
\be
P_\x{vap} = (E_\x{vap} + k_B T) \phi_\x{vap}.
\ee
In our calculations, we neglected the $k_B T$ term, which however does not have a significant impact on the results as it is much smaller than $E_\x{vap}$.

A simple RC circuit is used to model the coupling of the arc to the external power source. Although such a coupling system can be very complex \cite{paspow}, we use the simple RC approach adopted in the ArcPIC model \cite{arcpic}. The circuit has resistance $R$ and capacitance $C$ across the gap. To calculate the gap voltage and total current, we use the approach adopted in the ArcPIC model\cite{arcpic}
\begin{align}
  V_\x{gap} = V_0 + \frac{1}{C_\x{gap}} \int_0^t (I_\x{tot} - I_\x{gap}) \ \D t, \\
  I_\x{tot} = \frac{V_0-V_\x{gap}}{R},
\end{align}
where $V_\x{gap}$ is the gap voltage, $V_0$ is the applied voltage, $I_\x{gap}$ is the current moving through the gap and $I_\x{tot}$ is the total current in the circuit. The circuit has resistance $R$ and capacitance $C_\x{gap}$ across the gap as in \cite{arcpic}. The current induced by charges moving near the cathode is calculated using the \textit{Shockley-Ramo theorem} \cite{zhosr} as
\be
I = q \vb v \cdot \vb E_0(\vb r),
\ee
where $\vb v$ is velocity of the charged particle and $\vb E_0(\vb r)$ is the Laplace field at the position of the charge $\vb r$, i.e. the field that would be present without any charges and with the anode at unit potential. The gap current $I_\x{gap}$ is the sum of the currents induced by the electrons and ions $I_\x{gap} = \sum_k I_{\el}^k + \sum_k I_{\cu^+}^k + \cdots$.

\newcommand{\coulomb}{\parbox{2cm}{
\begin{align*}
\el + \el &\rightarrow \el + \el \\
\cu^+ + \el &\rightarrow \cu^+ + \el \\
\cu^+ + \cu^+ &\rightarrow \cu^+ + \cu^+ \\
&\vdots
\end{align*}}
}
\newcommand{\elastic}{\parbox{2cm}{
\begin{align*}
\cu + \el &\rightarrow \cu + \el \\
\cu + \cu &\rightarrow \cu + \cu
\end{align*}}
}
\newcommand{\ionization}{\parbox{2cm}{
\begin{align*}
\cu + \el &\rightarrow \cu^+ + 2 \el \\
\cu + \el &\rightarrow \cu^{2+} + 3 \el \\
\cu^+ + \el &\rightarrow \cu^{2+} + 2 \el
\end{align*}}
}
\newcommand{\exchange}{\parbox{2cm}{
\begin{align*}
\cu^+ + \cu &\rightarrow \cu + \cu^+
\end{align*}}
}
\newcommand{\recombination}{\parbox{2cm}{
\begin{align*}
\cu^+ + \el &\rightarrow \cu
\end{align*}}
}

\begin{table}[H]
\small
\centering
\resizebox{\columnwidth}{!}{\begin{tabular}{|c|c|c|}
\hline
  \multicolumn{2}{|c|}{Coulomb \cite{takcol}} \\
  \hline
  \multicolumn{2}{|c|}{
  \coulomb
}\\
  \hline
  Elastic \cite{matcol, tracs, arcpic} & Ionization \cite{matcol, bolcs, frecs, pincs, grecs} \\
  \hline
  \elastic & \ionization \\
\hline
  Charge exchange \cite{matcol, aubcs} & Recombination \cite{alics} \\
  \hline
  \exchange & \recombination \\
\hline
\end{tabular}}
\caption{Summary of collision processes.\label{tab:colls}}
\end{table}

The different particle species in our system are $\el$, $\cu$, $\cu^+$ and $\cu^{2+}$. These particle species can undergo collisions that are summarized in table \ref{tab:colls}. These include Coulomb collisions between all charged particles \cite{takcol}, elastic collisions \cite{matcol, tracs, arcpic, birvhs} between neutrals as well as neutrals and electrons, and impact ionization collisions \cite{matcol, bolcs, frecs, pincs, grecs} for neutrals and ions up to $\cu^{2+}$. In the present work, we only consider two-body interactions, but three-body interactions would be an interesting direction for future work. Since the cross sections of the collisions that lead to double ionization of Cu atoms are fairly small, we expect the formation of higher-order ions to be of low significance for plasma buildup, while single ionization of Cu atoms would contribute the majority of the ions. Charge exchange \cite{matcol, aubcs} and recombination \cite{alics} collisions can occur between $\cu$ and $\cu^+$, but these are also expected to be of lower significance as they do not produce new ions. A detailed description of the collisions implemented in FEMOCS is given in \cite{koimsc}.

\subsubsection{Dynamic weighting}\label{section:dynwei}

\begin{minipage}{\linewidth}
\dualfigc{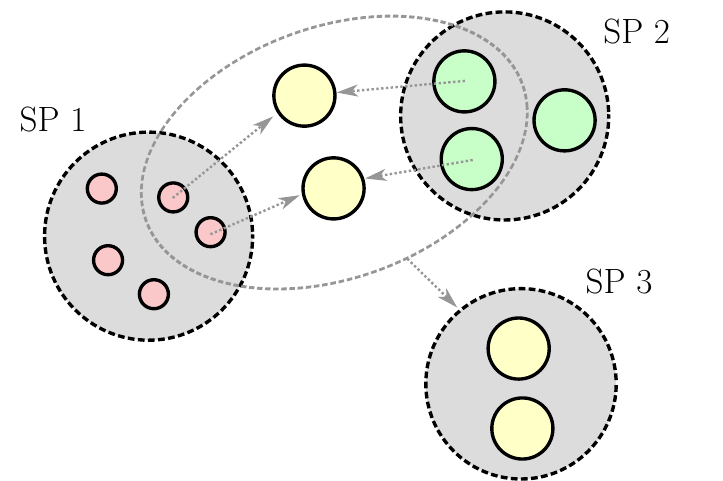}{Splitting of superparticles.\label{fig:spsplit}}
{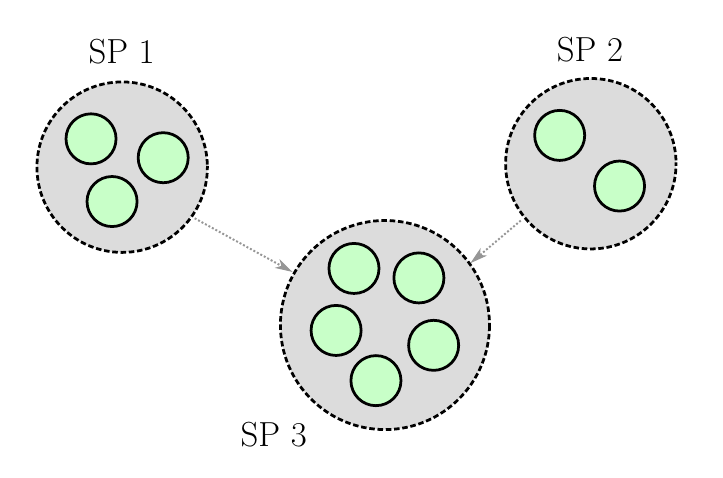}{Merging of superparticles.\label{fig:spmerge}}
{Dynamic weighting processes.}
\end{minipage}\medskip

By adjusting the weights of superparticles (splitting/merging), we can control the fidelity of a simulation \cite{lapdsw, teudsw, mardsw, gondsw}. In order for collisions to happen between superparticles that have different weights, we need to \textit{split} them when a collision occurs, in order to conserve the particle number \cite{koimsc}. The splitting process is illustrated in figure \ref{fig:spsplit}. Splitting of SPs means that the collision event occurring is no longer binary. Instead, we need to calculate the number of particles $N_\x{coll}$ that collide in each SP. This can be done by extending equation \ref{eq:colp} so that when two SPs with weights $w_1$ and $w_2$ collide, we calculate whether collisions happen for ``projectile'' sub-particles with weight $w_\x{p} = w_\x{min}$ and a single ``target'' superparticle with weight $w_t = \max(w_1, w_2)$.

The ``target'' is the SP with the larger weight, and the ``projectile'' sub-particles are split from the SP with the smaller weight. The number of sub-particles is $N_\x{p} = \min(w_1, w_2) / w_\x{min}$, which is an integer as $w_1$ and $w_2$ are multiples of $w_\x{min}$. The weight $w_\x{min}$ is the smallest allowable superparticle weight in the entire system, and SPs are injected into the system in multiples of $w_\x{min}$.

Collision probability for two colliding superparticles of different weights $w_a$ and $w_b$ is calculated as (see equation~\ref{eq:colp})
\be
  P(w_a, w_b) = 1 - \exp \lb -u n w_a w_b \sigma(E_0) \Delta t \rb, \\
\ee
where the cross section $\sigma(E_0)$ is calculated using the single particle ($w=1$) energy $E_0$. The collision cross section of the SPs with the weights $w_a$ and $w_b$ is then calculated as $w_a w_b \sigma(E_0)$.

The number of collisions $N_\x{coll}$ can be calculated by counting the number of sub-particle collisions occurring when a random number $U_i \sim \ud(0, 1)$ satisfies $U_i < P_i$
\be
N_\x{coll} = \sum_{i=1}^{N_\x{p}} \max \lb 0, \left\lceil 1 - \frac{U_i}{P(w_p^i, w_t)} \right\rceil \rb.
\ee
This would give the produced superparticle a weight of $N_\x{coll} w_\x{min}$.

\textit{Merging} is the reverse process to splitting. This means that we increase the weights of SPs in the system by combining two SPs into one \cite{koimsc}. This is shown in figure \ref{fig:spmerge}. This is beneficial for speeding up the simulations that have a large number of SPs. SPs of the same type can be combined together by adding their weights $w = w_1 + w_2$. The position and velocity of the new SP are
\begin{align}
r_m = \frac{r_1 \mu}{w_2} + \frac{r_2 \mu}{w_1}, \
v_m = \frac{v_1 \mu}{w_2} + \frac{v_2 \mu}{w_1},
\end{align}
where $\mu = \frac{w_1 w_2}{w_1 + w_2}$ is effective weight. The position of the new SP is the center of mass of the two smaller SPs. The merging process is an inelastic collision, which means it conserves momentum but does not conserve energy. This energy loss is typically sufficiently small to be neglected, which means it does not introduce significant errors.

When SPs are injected at the cathode surface, their weights can be adjusted based on the state of the simulation. For neutrals that evaporate from the surface, we determine the SP weights based on the number of evaporated atoms. Additionally, we adjust the SP weights to limit the SP number density in each cell. This means that even when the vapor flux and the number of particles in the system is large, the number of SPs remains manageable. A simplistic way to take both factors into account is to calculate the weights using the geometric mean
\be
w = \min \lb \left\lceil \max \lb 1, \sqrt{\frac{n_\x{cell}}{n_\x{max}} \frac{N_\x{inj}}{w_\x{min}}} \rb \right\rceil w_\x{min}, w_\x{max} \rb,
\ee
where $n_\x{cell}/n_\x{max}$ is a number density fraction in each cell above the surface of the cathode, $N_\x{inj}/w_\x{min}$ is the maximum number of injected SPs and $N_\x{inj}$ is the number of injected atoms. The number density fraction is determined from the number density of SPs in the cell $n_\x{cell} = N_\x{SP} / V_\x{cell}$ and some maximum SP number density $n_\x{max}$. In our simulations, we use $n_\x{max}=10^{-6}\usp{Å}{-3}$. When the number density in a cell is larger than the maximum, the injected SP weights will start to increase. Similarly, when the number of injected SPs increases, the weights increase. The $\min$, $\max$ and floor operations ensure that weights are multiples of the minimum SP weight $w_\x{min}$ and do not exceed the maximum SP weight $w_\x{max}$.

\subsubsection{Ion bombardment effects}

When ions impact the cathode surface, they can cause sputtering of neutral atoms from it. The number of ejected atoms is determined by the sputtering yield, which depends on the energy of the impacting ion. For Cu, we use experimentally fitted sputtering yields \cite{yamspuy} and energies \cite{stuspue} to determine the number and energy distribution of neutrals that are ejected from different parts of the cathode surface. The direction of the sputtered neutrals is assumed to have a uniform random distribution normal to the cathode surface, and their energy $E_s$ is sampled from the corresponding fitted distribution.

In addition to sputtering, ion bombardment of the cathode surface brings in kinetic energy which remains in the lattice in the form of surface heat, increasing the cathode temperature. The bombardment heat flux $P_\x{bom}$ on a face of area $A$ is
\be
P_\x{bom} = \frac{E_b - E_s}{A \Delta t_\x{heat}},
\ee
where $E_b$ is the total energy of the ions that have bombarded the surface during a heating time step $\Delta t_\x{heat}$ and $E_s$ is the corresponding total energy of sputtered neutrals.

\subsubsection{Direct field ionization}

\textit{Direct field ionization} is another way neutrals can be ionized, apart from impact ionization. When the local electric field is sufficiently high, electrons can directly escape from a neutral atom by quantum tunneling, similarly to the field emission process. It can be estimated that direct field ionization becomes more significant than impact ionization at electric fields of $F_\x{loc} \approx 8 \us{GV/m}$ \cite{caldfi}. This is why including direct field ionization in our simulations is important, as we expect it to dominate ionization processes at high fields. The Ammosov-Delone-Krainov (ADK) model gives us the probability of ionization per femtosecond as \cite{caldfi, brudfi}
\be
P = \frac{1.52 \times 4^n \xi}{n \Gamma(2 n) \us{fs}} \lb \frac{20.5 \xi^{3/2}}{E} \rb^{2 n - 1} \mkern-18mu \exp \lb \frac{-6.83 \xi^{3/2}}{E} \rb,\label{eq:dfi}
\ee
where $n = 3.69 z \xi^{-1/2}$, $\xi$ is the potential of ionization (eV), $E$ is the electric field (GV/m), $z$ is the charge number after ionization and $\Gamma(z)$ is the Gamma function. The ionization potential $\xi$ is the sum of ionization energies up to a certain ionization level.

The probability for a neutral to be ionized within a time step $\Delta t$ is $P \Delta t$. To determine whether ionization occurs, we use a random number $U \sim \ud(0, 1)$ with the criterion $P \Delta t > U$. We reject all neutrals in fields greater than the barrier suppression limit $E > \frac{3}{2} \xi^{3/2}$. Injection of ion SPs in direct field ionization occurs similarly as in impact ionization, with SP splitting if necessary.

\section{Results}

\subsection{Heating effects in a static nanotip}

\quadfigcd{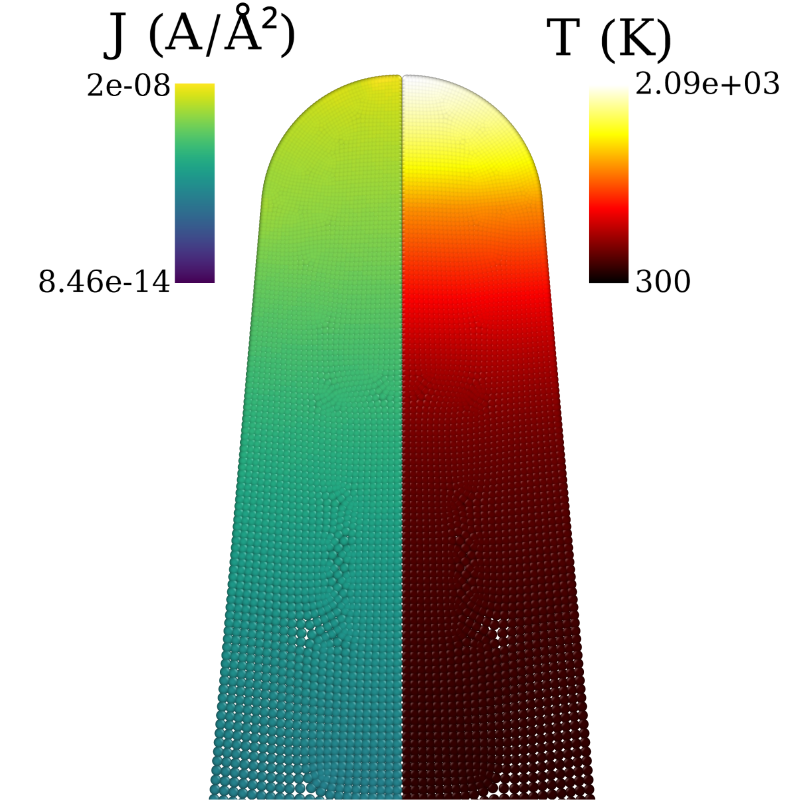}{$\gamma=5\degs, t=500 \us{ps}$.\label{fig:tdist5_500ps}}
{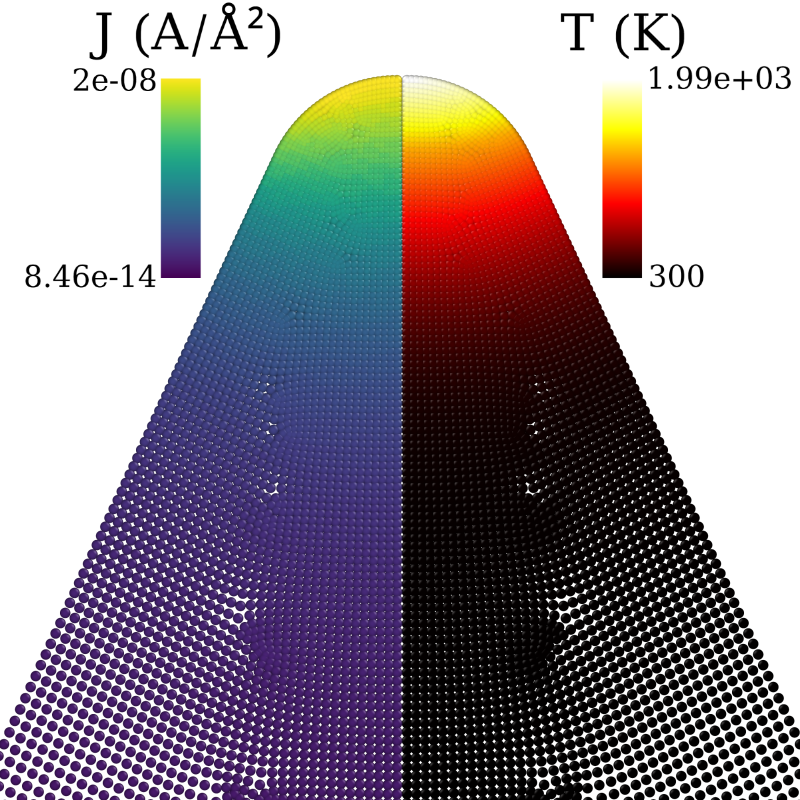}{$\gamma=25\degs, t=500 \us{ps}$.\label{fig:tdist25_500ps}}
{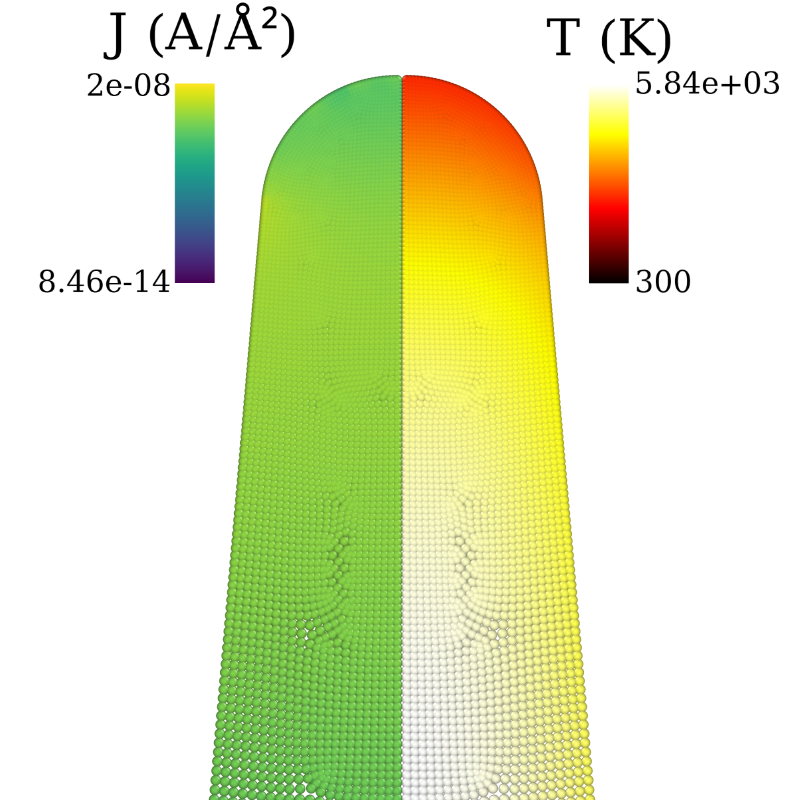}{$\gamma=5\degs, t=5 \us{ns}$.\label{fig:tdist5_5ns}}
{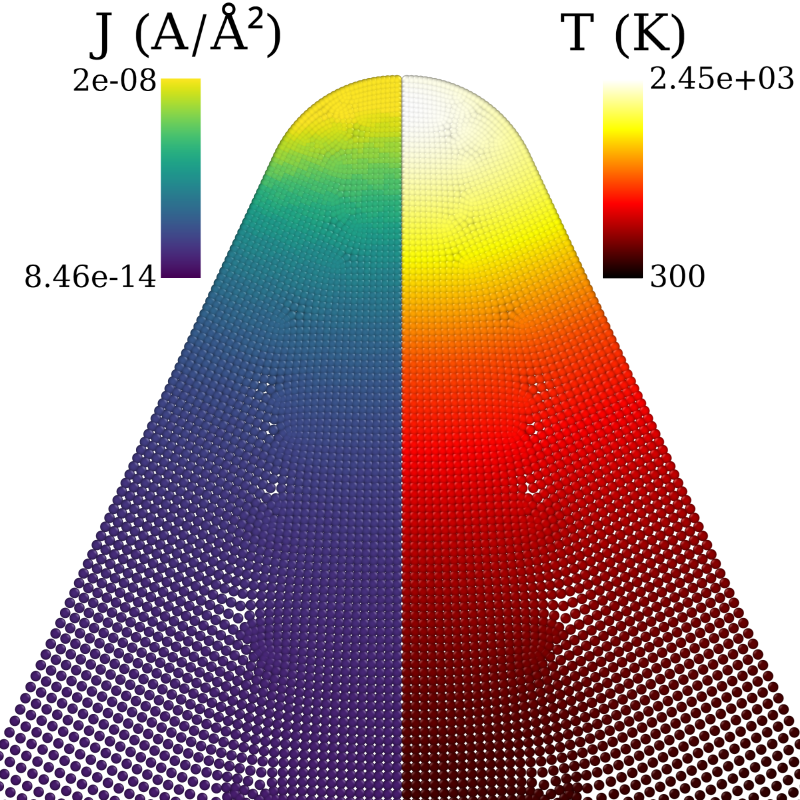}{$\gamma=25\degs, t=5 \us{ns}$.\label{fig:tdist25_5ns}}{Current density $J$ and temperature $T$ distributions (at times $t=500 \us{ps}$ and $t=5 \us{ns}$) at the tops of the nanotips with different tip opening angles $\gamma$.}

Since heating dynamics plays a crucial role in the runaway process, we first study how heating can be affected by the shape of a Cu nanotip.
In these simulations, we focus only on electron emission, while evaporation of neutral $\cu$ atoms is disabled. Properties of Cu (such as work function, resistivity and heat of vaporization) are the same in all of our simulations as those used in previous work \cite{femocs, getelec, kyrhea}. Emission current density on the cathode surface is given by $\vb J_e$ (equation \ref{eq:bce1}), calculated using GETELEC.
Two cases are simulated with different tip opening angles of $\gamma=5\degs$ and $\gamma=25\degs$ (figures \ref{fig:tdist5_500ps}--\ref{fig:tdist25_5ns}). The geometry of the nanotip is assumed to be static, i.e. it is not changing its shape with time. Because of the boundary conditions of the studied geometry, we expect that the opening angle of the nanotip will affect the current density distribution within it, and therefore the heat distribution. To determine this, we simulate two Cu nanotips with identical top radii of curvature $r=50\us{nm}$ of the same height $h=50r$, changing only the opening angle. The gap distance from the cathode bottom surface to the anode in both cases is $h_\x{gap}=150r$. The circuit has resistance $R=1\us{k\ensuremath{\Omega}}$ and capacitance $C=1 \us{pF}$. To ensure sufficiently strong field emission, each tip has a different voltage applied between the cathode and anode such that the local electric field is $F_\x{loc}=10 \us{GV/m}$ at the top of each tip. The tips with different opening angles resulted in different field enhancement factors $\beta=38.3$ and $\beta=29.1$ for the tips with $\gamma=5\degs$ and $\gamma=25\degs$, respectively.
In these simulations, the maximum temperature reached its stable value after $5 \us{ns}$ of simulation time.

Spatial current density and temperature distributions (bulk mesh nodes), obtained with the visualization tool OVITO\cite{ovito}, are shown in figures \ref{fig:tdist5_500ps} -- \ref{fig:tdist25_5ns}. Note that in these figures, in contrast to figure \ref{fig:domains}, only the top parts of the simulated nanotips are shown for better illustration.
Spatial distributions for both current density (left) and temperature (right) are shown in the same figure. The current density distributions are time-averaged over $250 \us{ps}$ to remove rapid fluctuations caused by shot noise due to electron emission.

We might expect that the heating of the nanotip will take place at the top of it, where the emission current is highest due to the maximum electric field. However, in the temperature distributions shown in figures \ref{fig:tdist5_5ns} and \ref{fig:tdist25_5ns}, we can see that the temperature maximum shifts down for the narrow tip, while it remains at the top for the wide one. Two factors contribute to the shift in figure \ref{fig:tdist5_5ns}: the current density distribution inside of the nanotip (which determines the amount of Joule heating \ref{eq:heat}) and Nottingham cooling at the top of the nanotip.

Current density is dependent on the total emission current and cross-sectional area of the nanotip. For the narrow tip \ref{fig:tdist5_5ns}, the cross-sectional area is smaller than for the wide tip \ref{fig:tdist25_5ns}. As a consequence of the continuity equation \ref{eq:cont}, current emitted at the top of the nanotip must travel through the entire length of the nanotip. This means that for the narrow tip, current densities throughout the tip will be higher (for the same emission current). Joule heating also increases due to the resistivity of $\cu$, which increases with temperature. In the emission region at the top, total current through a section of the nanotip decreases due to the emitted current. 

Initially, when the nanotip is cold (see figures \ref{fig:tdist5_500ps} and \ref{fig:tdist25_500ps}), the emission current is determined only by field emission, which is strongest at the top of the nanotip. Hence, for both geometries, the maximum current density is located at the tops of both tips. High current density at the top causes Joule heating that heats up the material at the emitting spot and the thermal field emission gradually transforms to be more dominated by thermionic emission. Heating of the emitting spot leads to the spread of the emission area, which increases the emission current. Joule heating for the narrow tip \ref{fig:tdist5_5ns} is greater than for the wide tip \ref{fig:tdist25_5ns} because of the difference in cross-sectional area. This results in stronger heating of a larger part of the narrow tip (down from the top), while for the wide tip the heat is mostly concentrated at the top of it. The wide tip also has stronger heat dissipation at the bottom of the tip because of larger surface area.

\dualfigcd{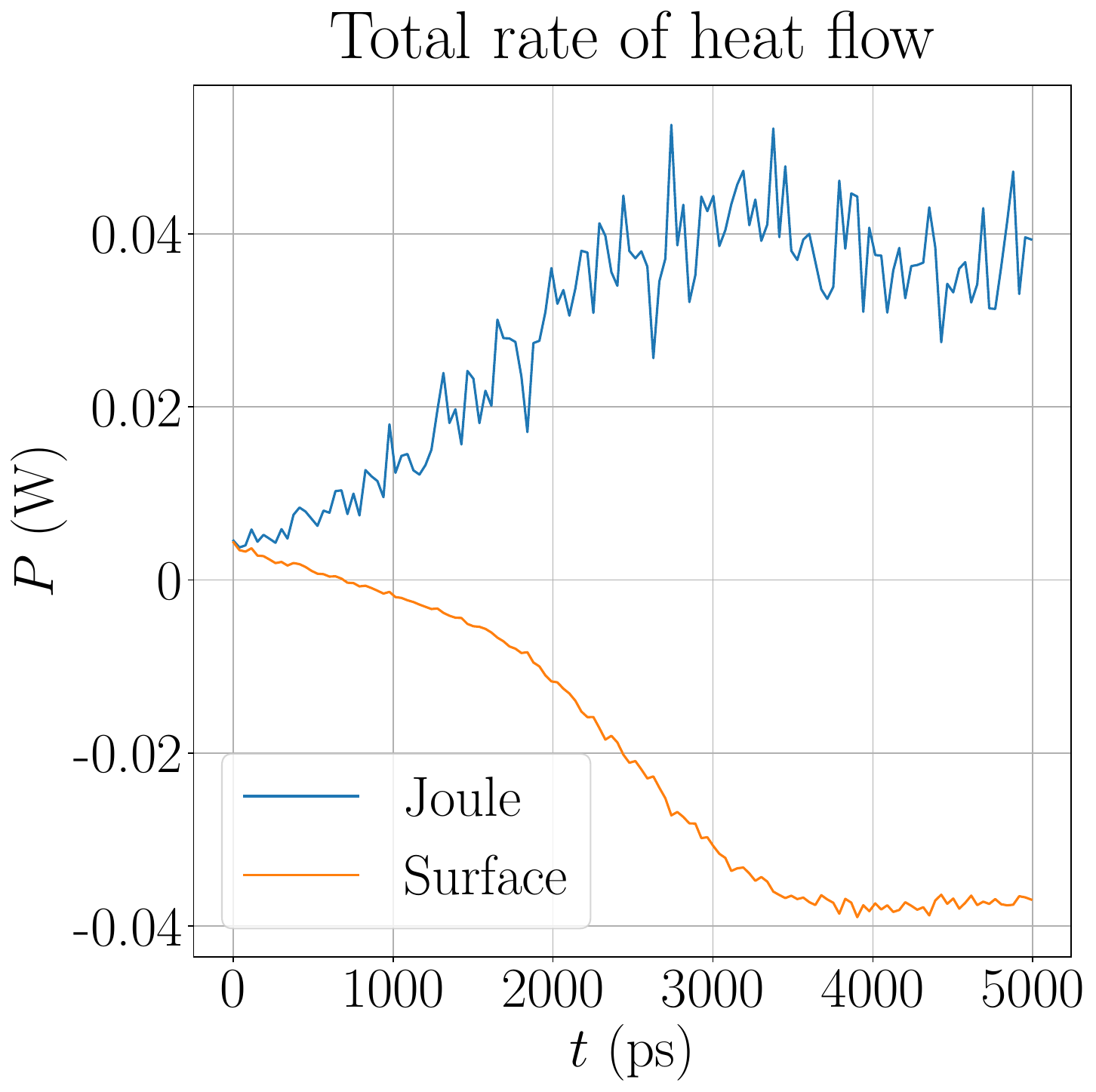}{$\gamma=5\degs$\label{fig:bulk5}.}
{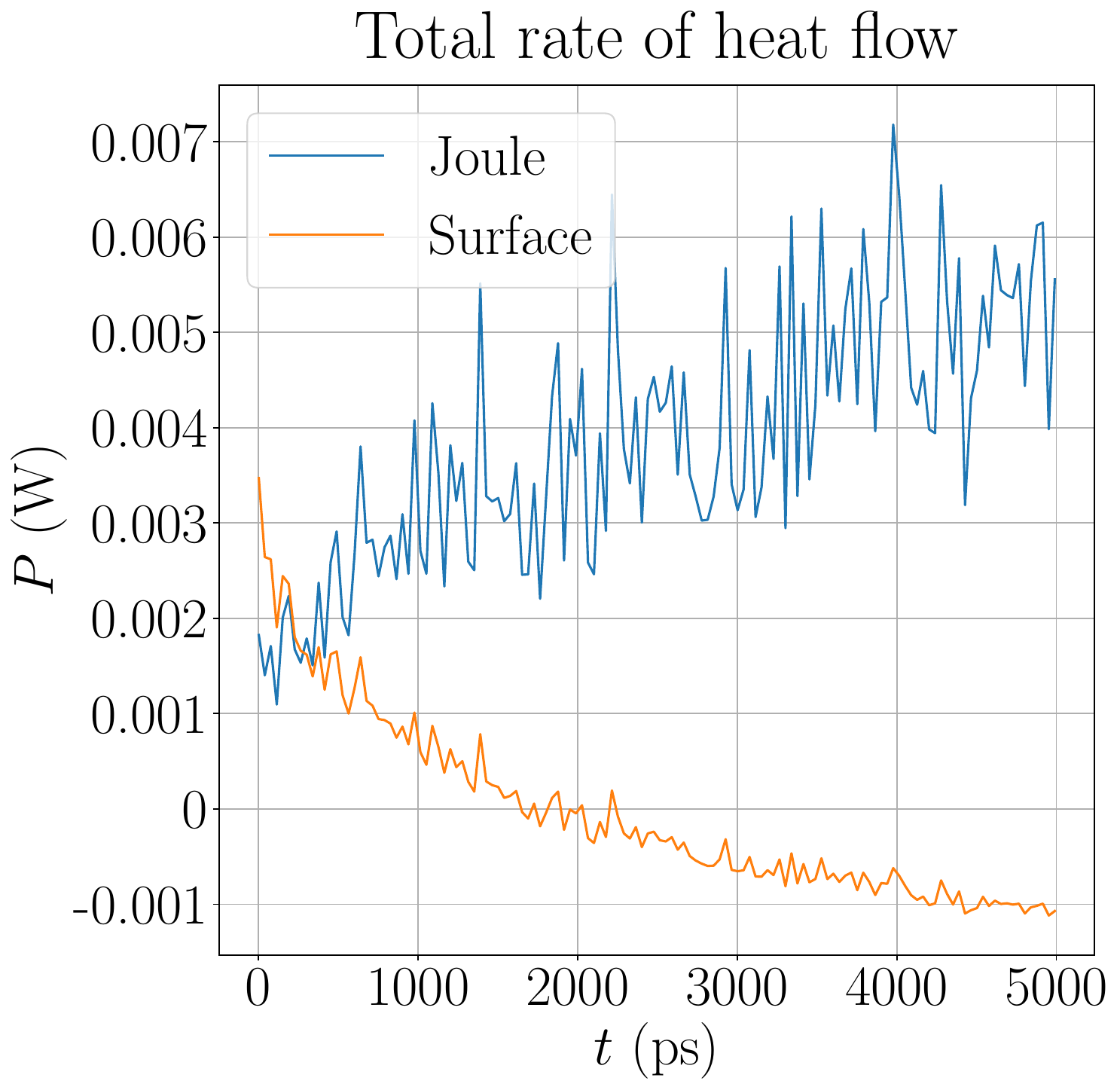}{$\gamma=25\degs$\label{fig:bulk25}.}
{Total rate of heat flow in the cathode bulk as a function of time generated from different sources for nanotips with different opening angles $\gamma$.}

At high temperatures, the Nottingham effect contributes to cooling of the cathode surface where strong electron emission occurs. The resulting temperature distribution is determined by the combination of the Nottingham effect on the surface (cooling at high temperatures) and Joule heating. The Nottingham cooling at the top of the tip
reduces thermionic emission, which in turn reduces Joule heating. The total Joule heat and the Nottingham heat (flow rate) are compared for the narrow and the wide tips in figures \ref{fig:bulk5} and \ref{fig:bulk25}. We see that the Joule heating in the narrow tip is stronger and the cooling on its surface via the Nottingham effect is more efficient. For the wide tip, Joule heating dominates over the Nottingham cooling, as the net rate of heat flow (Joule + Nottingham) is greater for the wide tip compared to the narrow one.

In both cases the temperature is well above the melting point of copper. Since melting causes a phase transition from solid to liquid, the material response to the electric field and hence the temperature distribution will be different in the dynamic system compared to a static tip. However, the results of the present simulations can provide sufficient insight on the role of different processes in the initiation of plasma onset.

\subsection{Runaway and plasma onset}

\trifighs{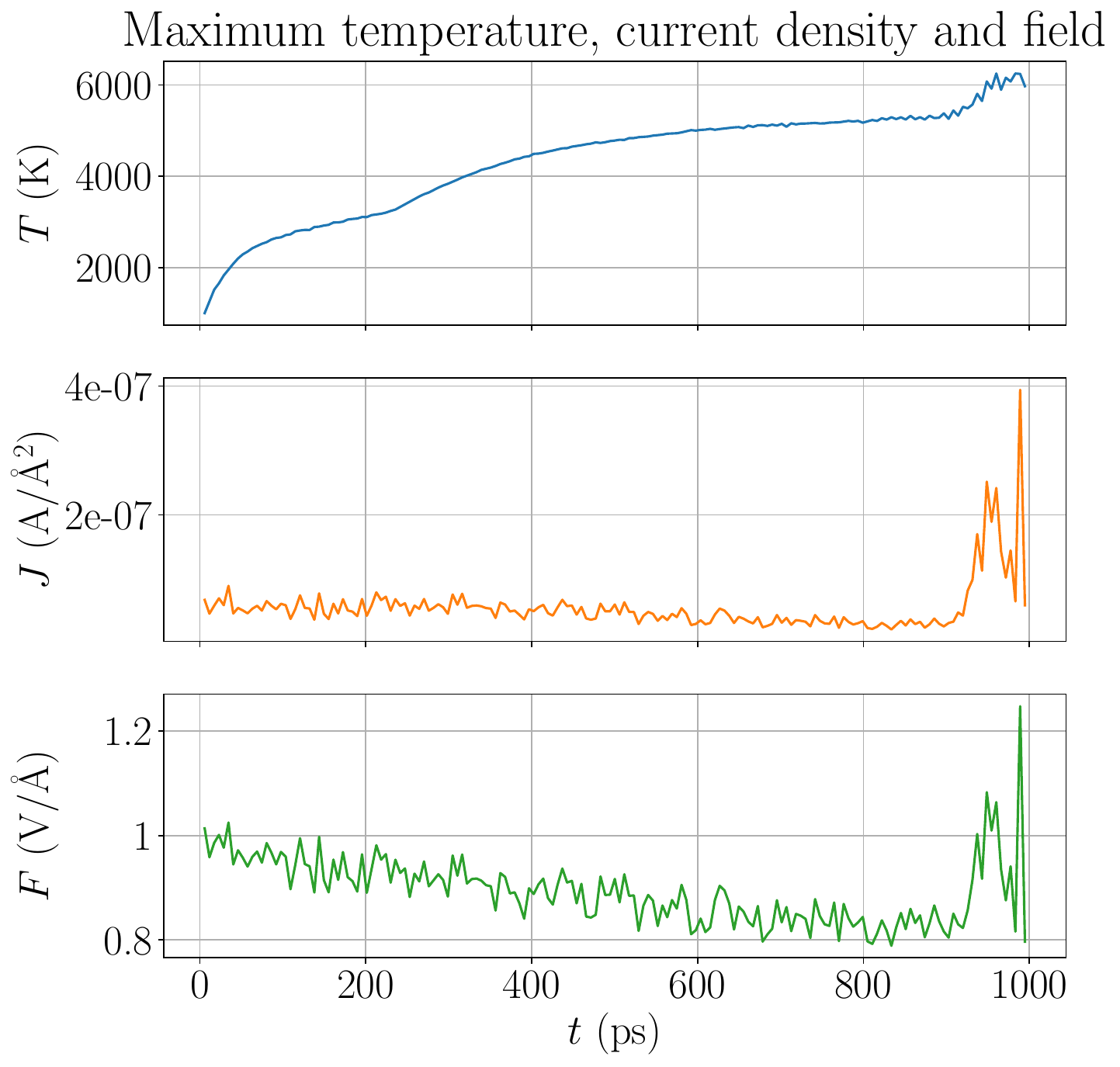}{\footnotesize Cathode surface maximums as a function of time (temperature $T$, current density $J$ and field $F$). \label{fig:tjf13}}
{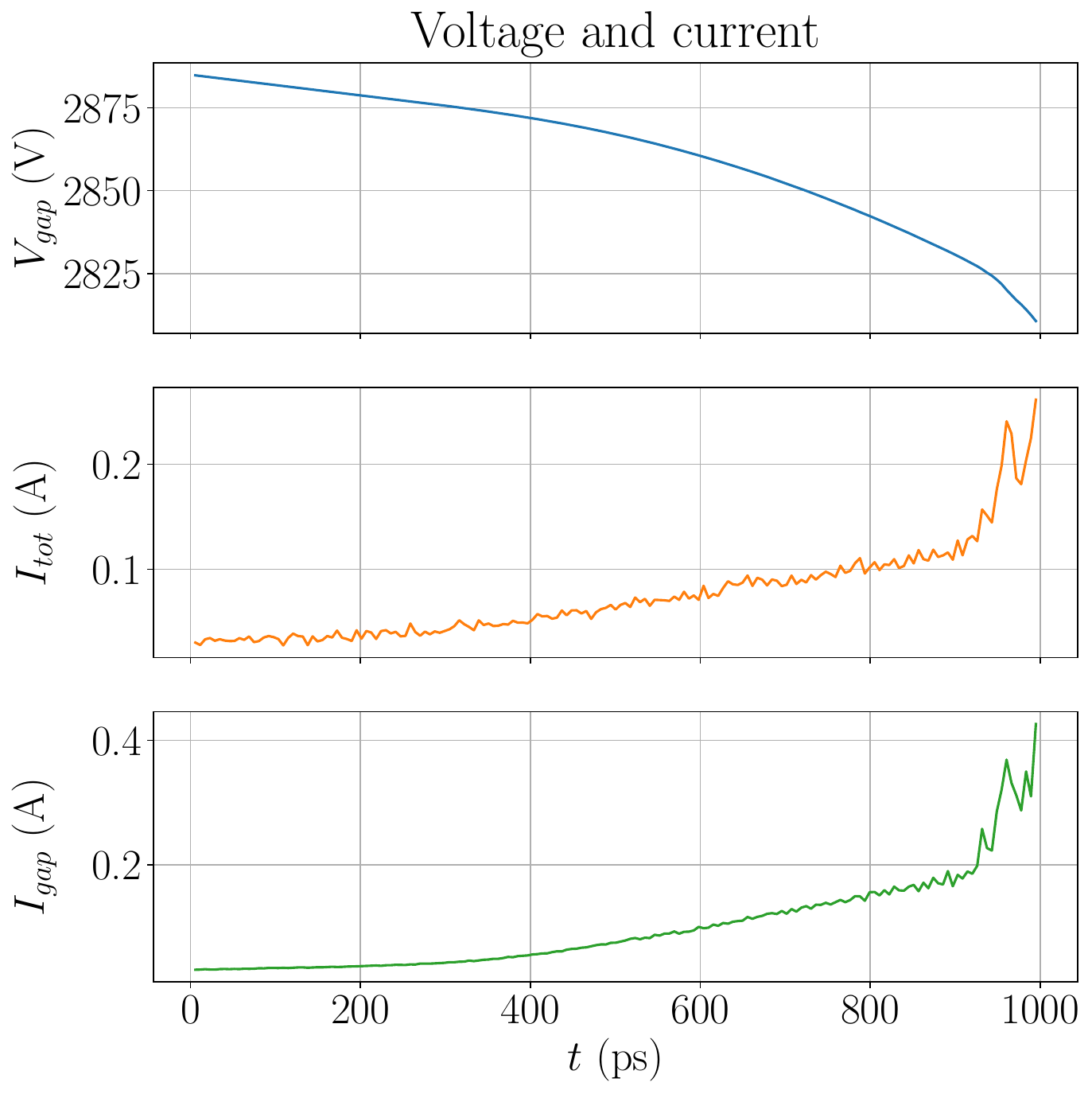}{\footnotesize Gap voltage $V_\x{gap}$, total circuit current $I_\x{tot}$ and gap current $I_\x{gap}$ as a function of time.\label{fig:vi13}}
{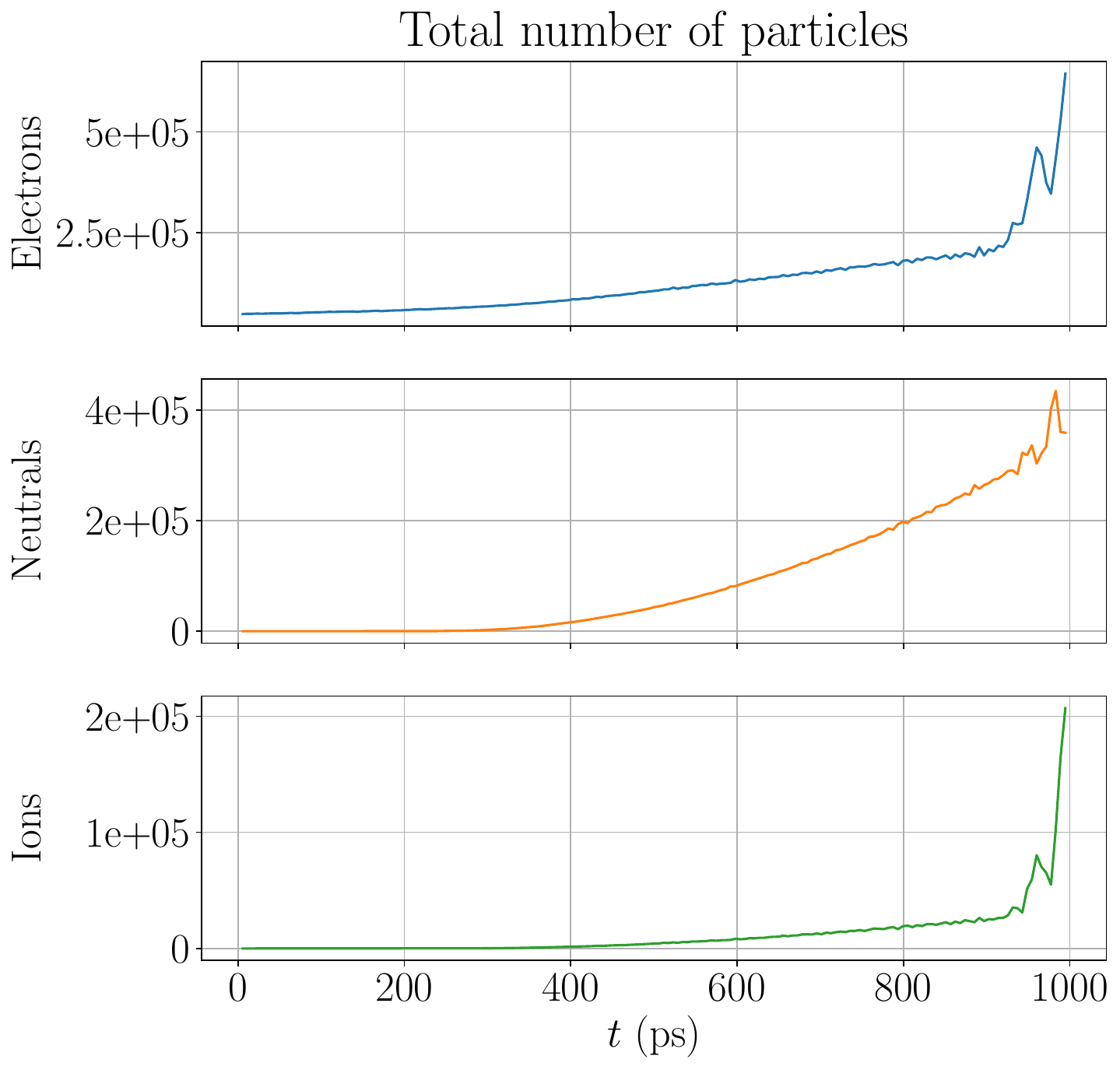}{\footnotesize Total number of different particles in the system (electrons, neutrals and ions) as a function of time.\label{fig:nps13}}
{System state for the tip with a local field of $F_\x{loc} = 13 \us{GV/m}$.}{0.95}
\trifighs{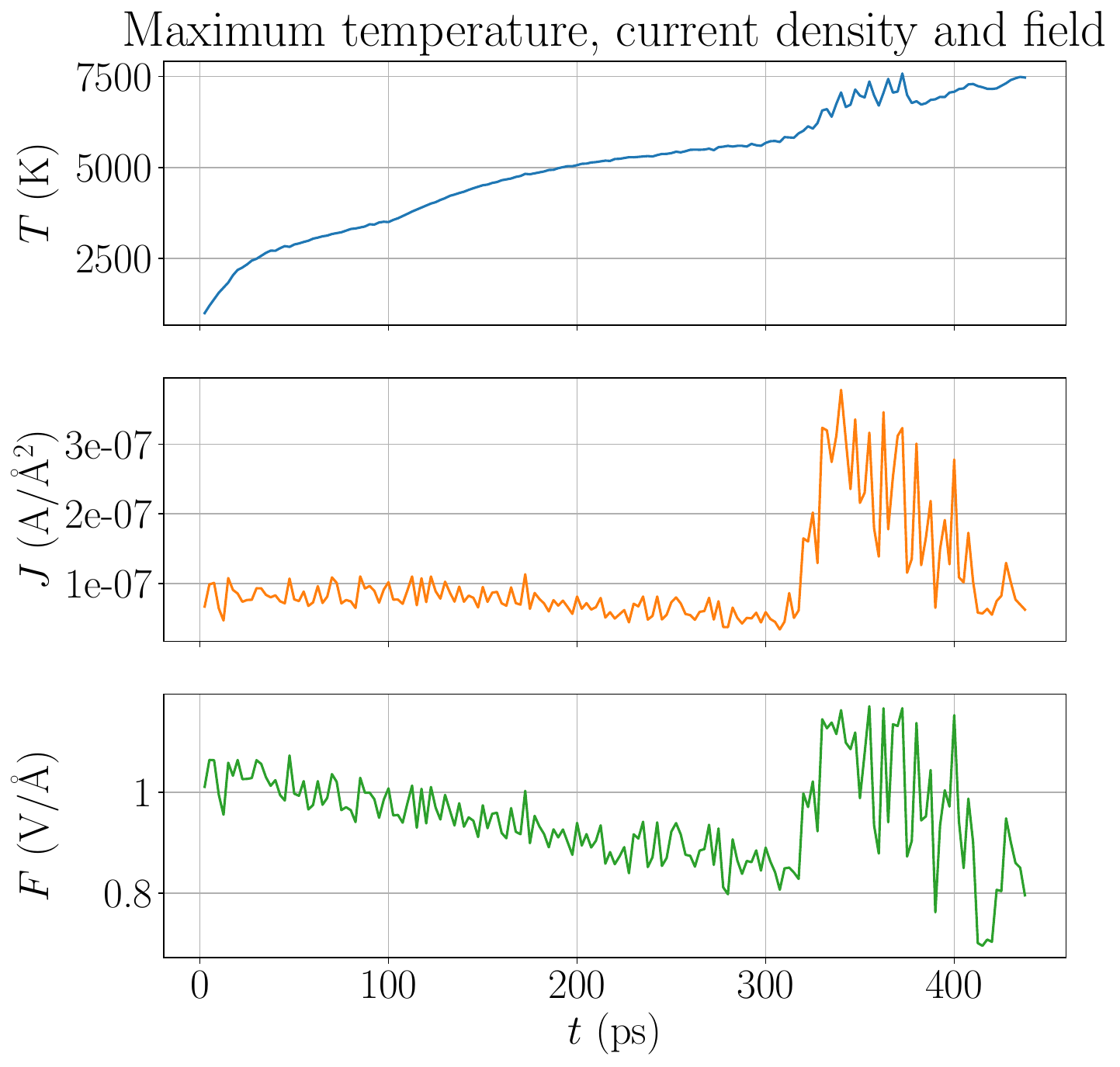}{\footnotesize Cathode surface maximums as a function of time (temperature $T$, current density $J$ and field $F$).\label{fig:tjf15}}
{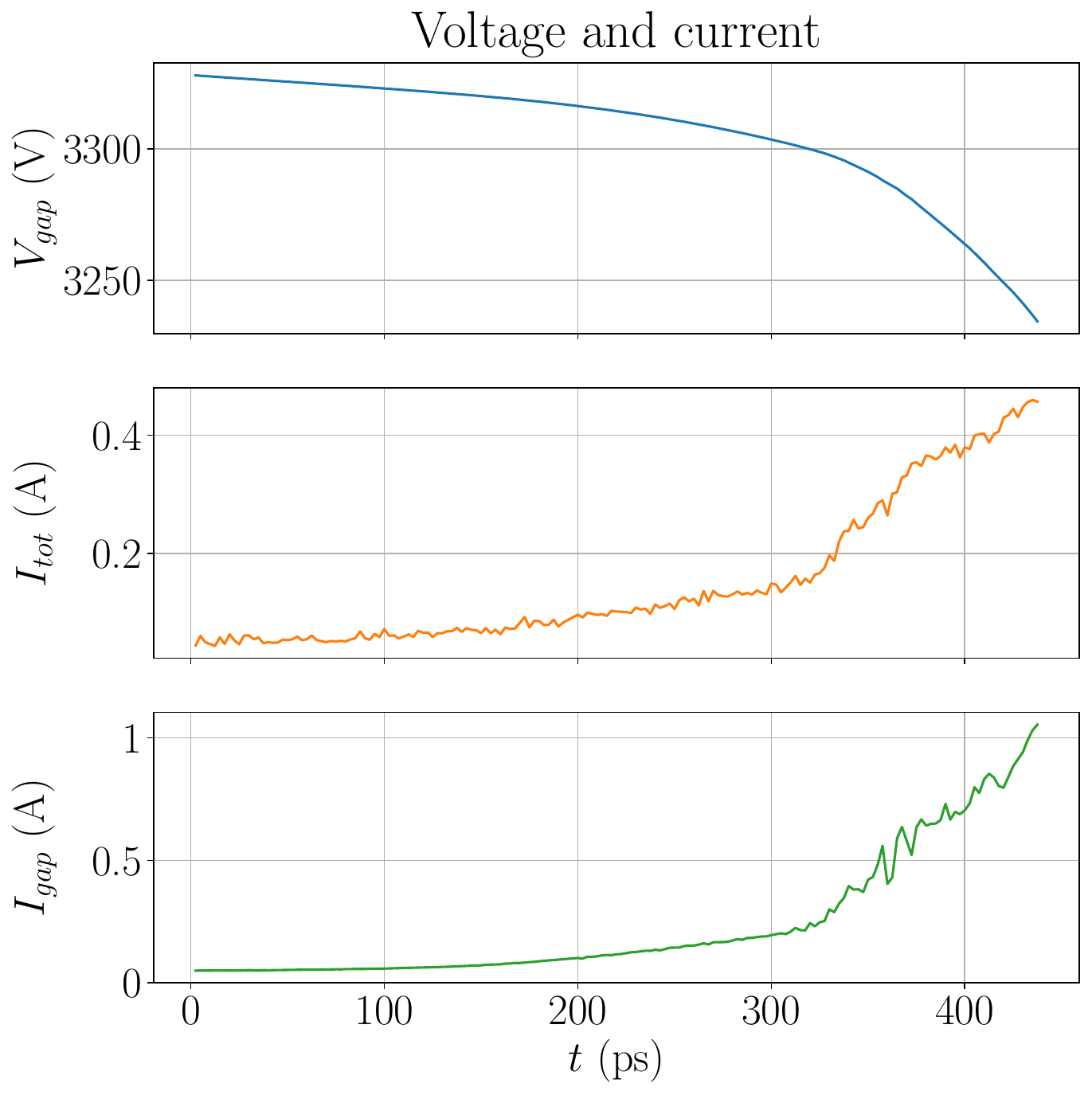}{\footnotesize Gap voltage $V_\x{gap}$, total circuit current $I_\x{tot}$ and gap current $I_\x{gap}$ as a function of time.\label{fig:vi15}}
{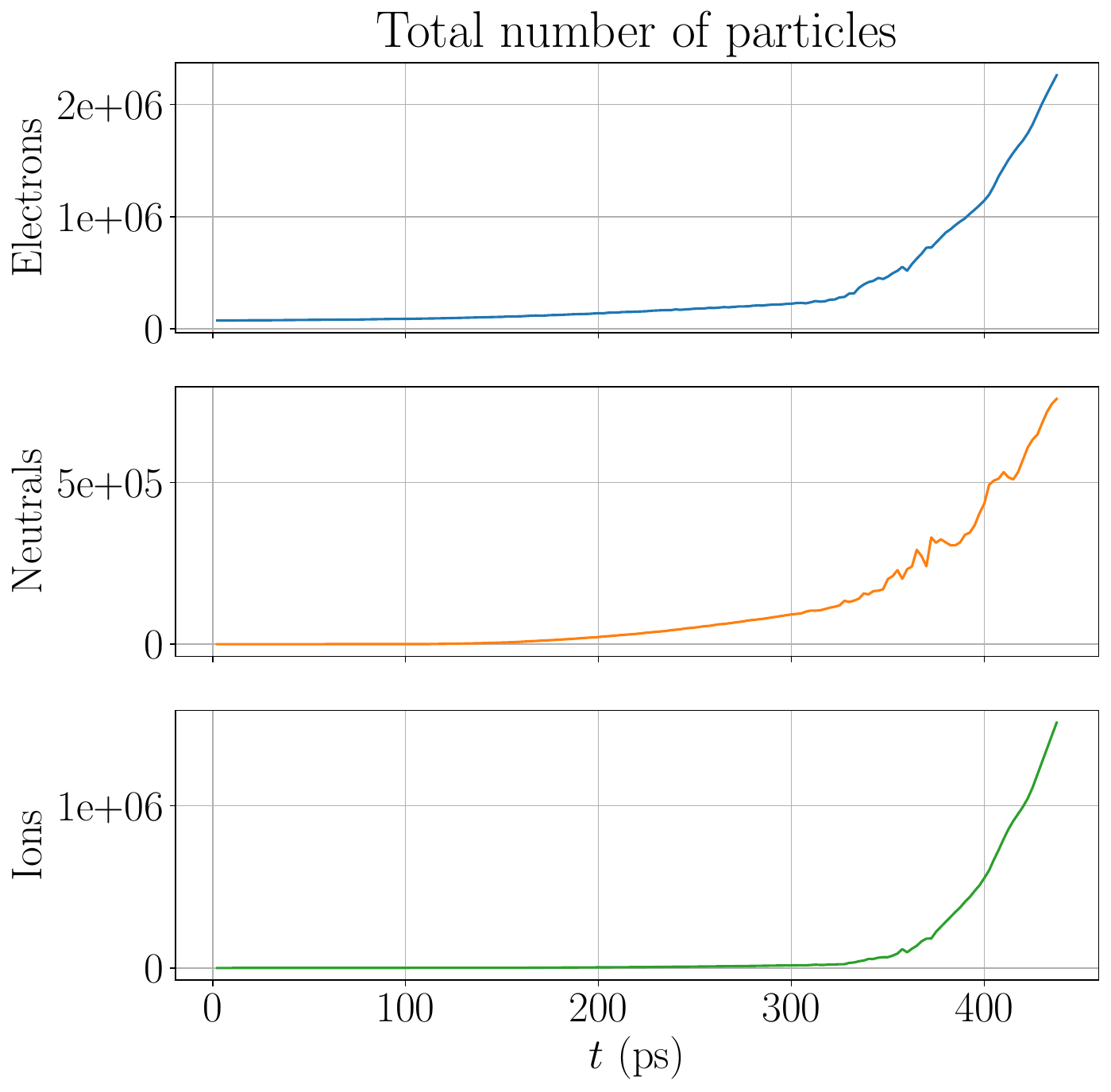}{\footnotesize Total number of different particles in the system (electrons, neutrals and ions) as a function of time.\label{fig:nps15}}
{System state for the tip with a local field of $F_\x{loc} = 15 \us{GV/m}$.}{0.95}
\trifighs{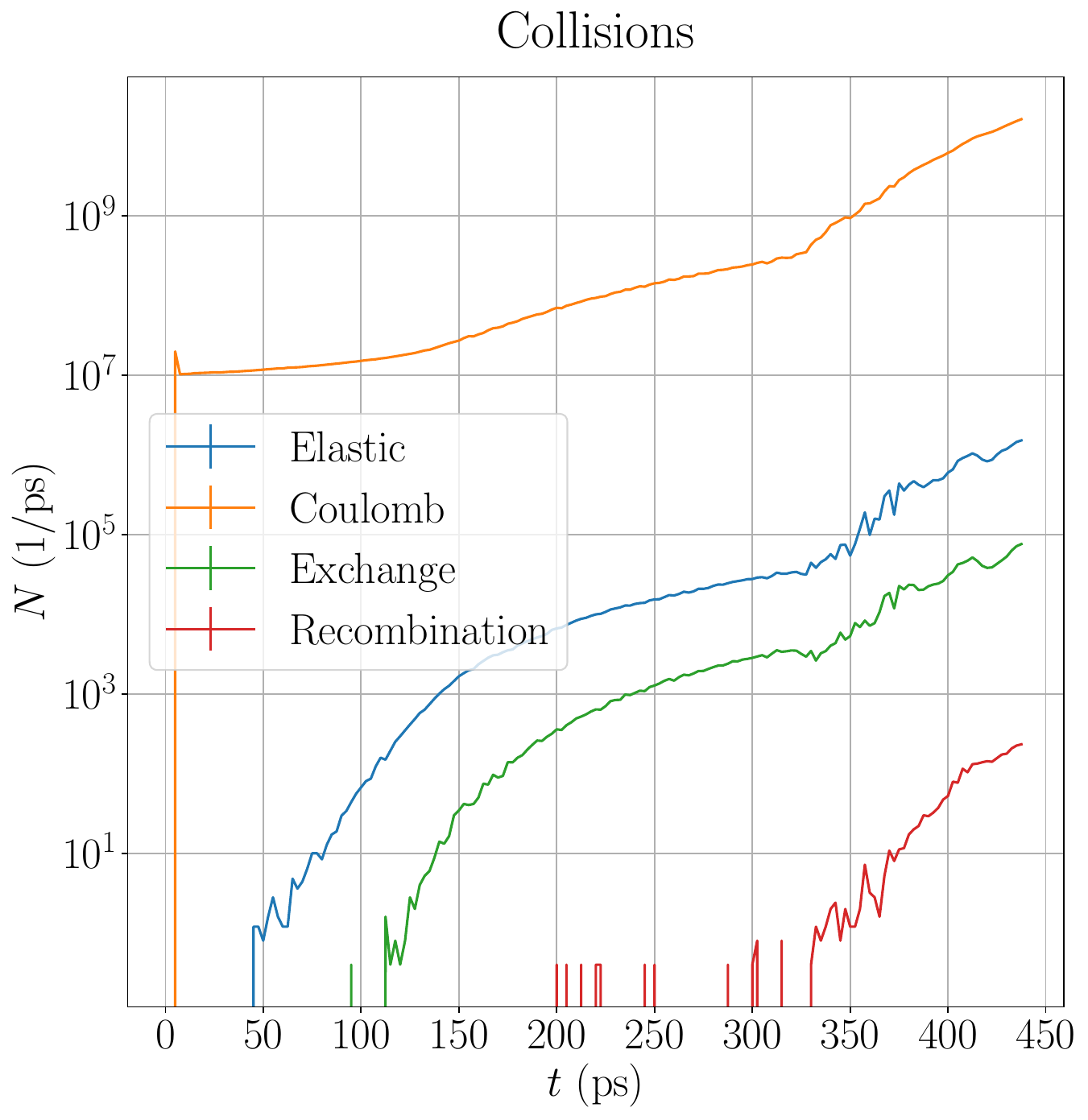}{\footnotesize Frequency of elastic, Coulomb, exchange and recombination collisions as a function of time.\label{fig:elastic}}
{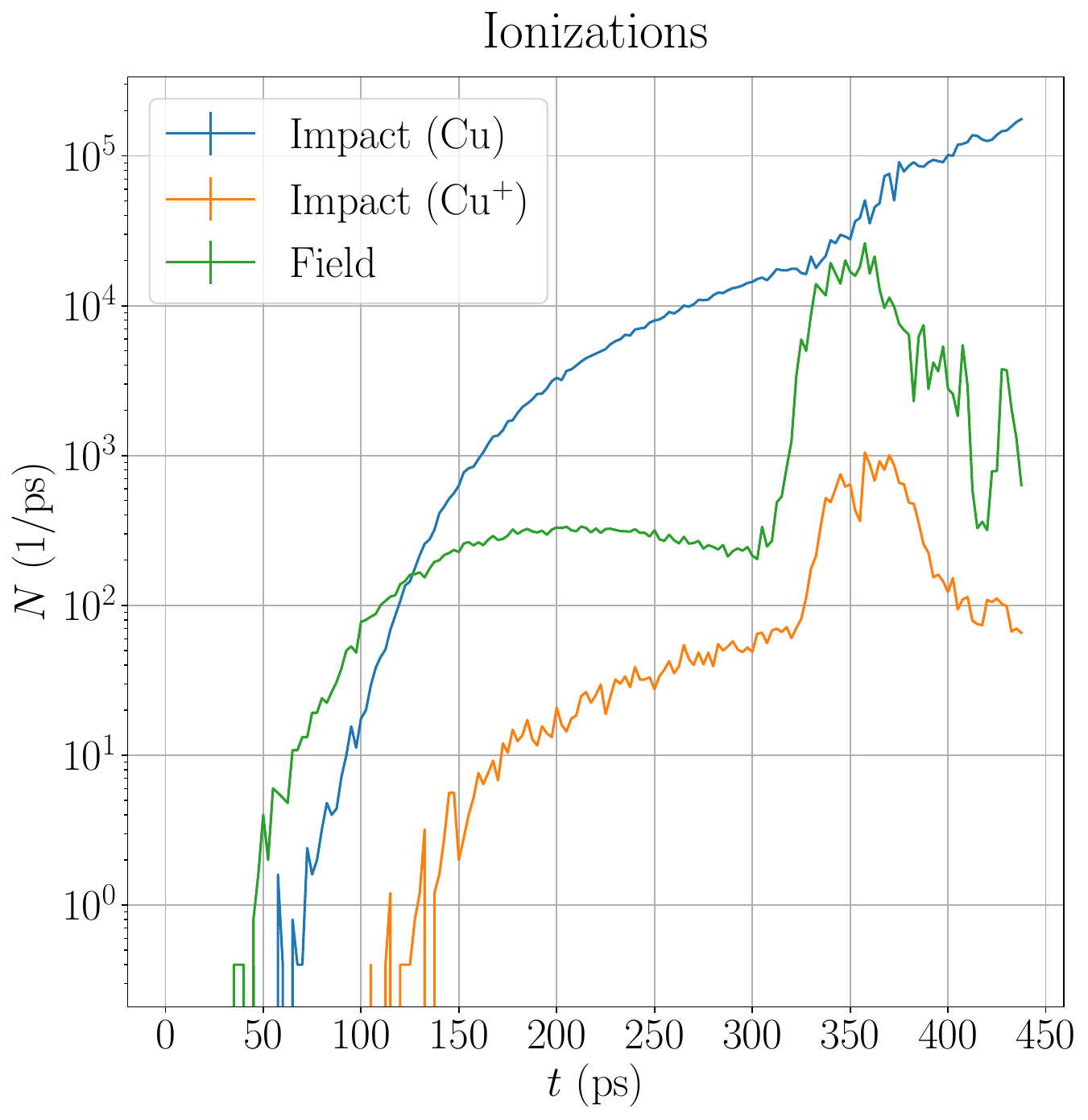}{\footnotesize Frequency of impact ionizations (Cu, Cu$^+$) and field ionizations as a function of time.\label{fig:ionization}}
{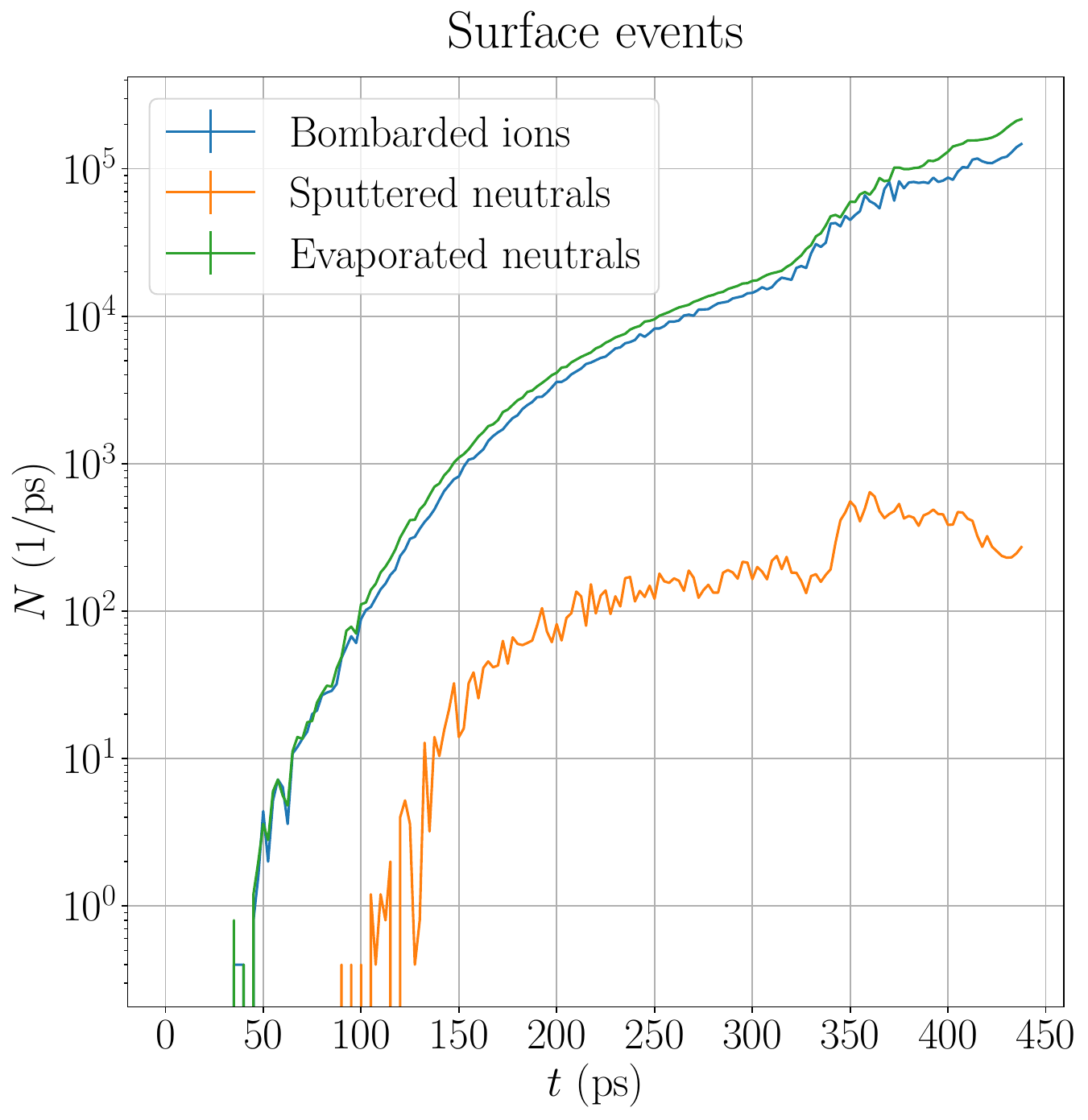}{\footnotesize Frequency of cathode surface interactions as a function of time.\label{fig:neutrals}}
{Particle interaction events for the tip with a local field of $F_\x{loc} = 15 \us{GV/m}$.}
{0.95}

The initial development of plasma around a nanotip was simulated to study the effect of surface interactions on the runaway process. We simulated a static nanotip at different local electric fields $F_\x{loc}$ of $10 \us{GV/m}$, $13 \us{GV/m}$ and $15 \us{GV/m}$ to determine when thermal runaway occurs. The nanotip has dimensions $r=50 \us{nm}$ and $h=50r$ similarly to the previous section, but with tip angle $\gamma=15\degs$. The gap distance is $h_\x{gap}=150r$. The simulation box has dimensions $w_\x{box} = h_\x{gap} / 2$ and $h_\x{box} = h/2 + h_\x{gap}$. The simulation domains were discretized into a quadrilateral mesh that contains 13449 elements for $\Omega_\x{bulk}$ and 18305 elements for $\Omega_\x{vac}$. The element area ranges from about $100\usp{Å}{2}$ to $10^6 \usp{Å}{2}$. A time step of $0.1\us{fs}$ is used for PIC and $50\us{fs}$ for heat calculations. The circuit has resistance $R=1\us{k\ensuremath{\Omega}}$ and capacitance $C=1 \us{pF}$. In this simulation, we enabled all of the interactions described in the methods section. Neutrals evaporate from the cathode as a result of heating. The PIC collisions in this simulation are elastic, Coulomb, ionization, exchange and recombination. For these simulations, SP splitting was enabled and merging disabled. Additionally, field ionization, sputtering and bombardment heating are all enabled. The simulation time for most of these runs ranged from a few days up to a week. Particle plasma simulation required the most computational time, as the number of particles grows very large. The particle simulations are parallelized using multithreading (16 threads for these runs).

For the simulation with $F_\x{loc} = 10 \us{GV/m}$, we do not see significant plasma formation, and it appears that the cathode is not going into thermal runaway. The state of the system is shown in figures \ref{fig:tjf13}--\ref{fig:nps13} for $F_\x{loc} = 13 \us{GV/m}$ and in figures \ref{fig:tjf15}--\ref{fig:nps15} for $F_\x{loc} = 15 \us{GV/m}$. Figures \ref{fig:tjf13} and \ref{fig:tjf15} show the maximum values for temperature, emission current and field on the surface of the nanotip as a function of time. Figures \ref{fig:vi13} and \ref{fig:vi15} show the voltage and current evolution of the system. In figures \ref{fig:nps13} and \ref{fig:nps15}, we show the total number of particles of each type in the system (electrons, neutrals and ions). The number of ions is the sum of each ion species ($N_{\x{Cu}^+} + N_{\x{Cu}^{2+}}$). The frequencies of particle interaction events are shown in figures \ref{fig:elastic}--\ref{fig:neutrals}.

For a local field of $13 \us{GV/m}$, the temperature increases smoothly before reaching $5000 \us{K}$. After $900 \us{ps}$, a rapid increase in temperature can be observed. For a local field of $15 \us{GV/m}$, there is a similar sudden spike in temperature after $300 \us{ps}$, with corresponding increases in emission current and field. At the same time, voltage is seen decreasing rapidly with an increase in circuit current. It is at this point that we observe a large number of ionization events in figure \ref{fig:ionization}. This shows that evaporated neutrals above the surface are ionizing, forming a cloud of plasma with a neutral charge. At the start of plasma formation, field ionizations are shown to be more frequent than impact ionizations. When the plasma has formed, this is reversed as the field in the plasma decreases dramatically. We can also see that $\cu^+$ ions are more important for the formation of the plasma, since double ionizations are relatively rare at this stage of plasma development. In figure \ref{fig:neutrals}, we show events occurring on the surface of the cathode, such as ion bombardments (contributing to heating), sputtering and evaporation. It can be seen that neutral evaporation is much more important than sputtering.

\fig{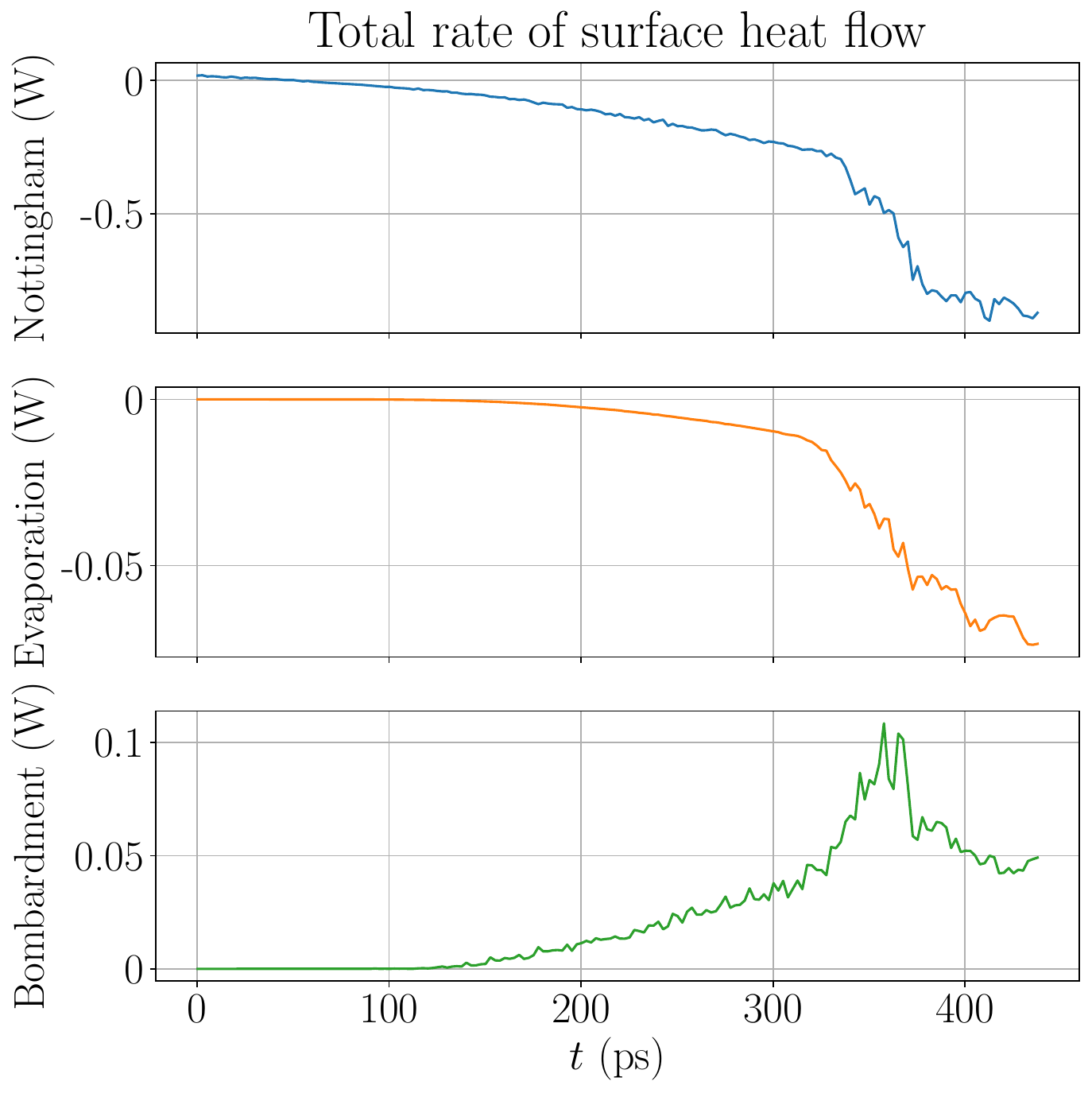}{Total rate of heat flow on the cathode surface as a function of time generated from different sources at a local field of $F_\x{loc} = 15 \us{GV/m}$. \label{fig:thsem15}}{0.7}

Heat sources on the cathode surface are compared in figure \ref{fig:thsem15}. We can see that Nottingham heat accounts for most of the rate of heat flow on the surface, followed by evaporation heat and bombardment heat. Bombardment heat is approximately as significant as evaporation heat.

\trifighs{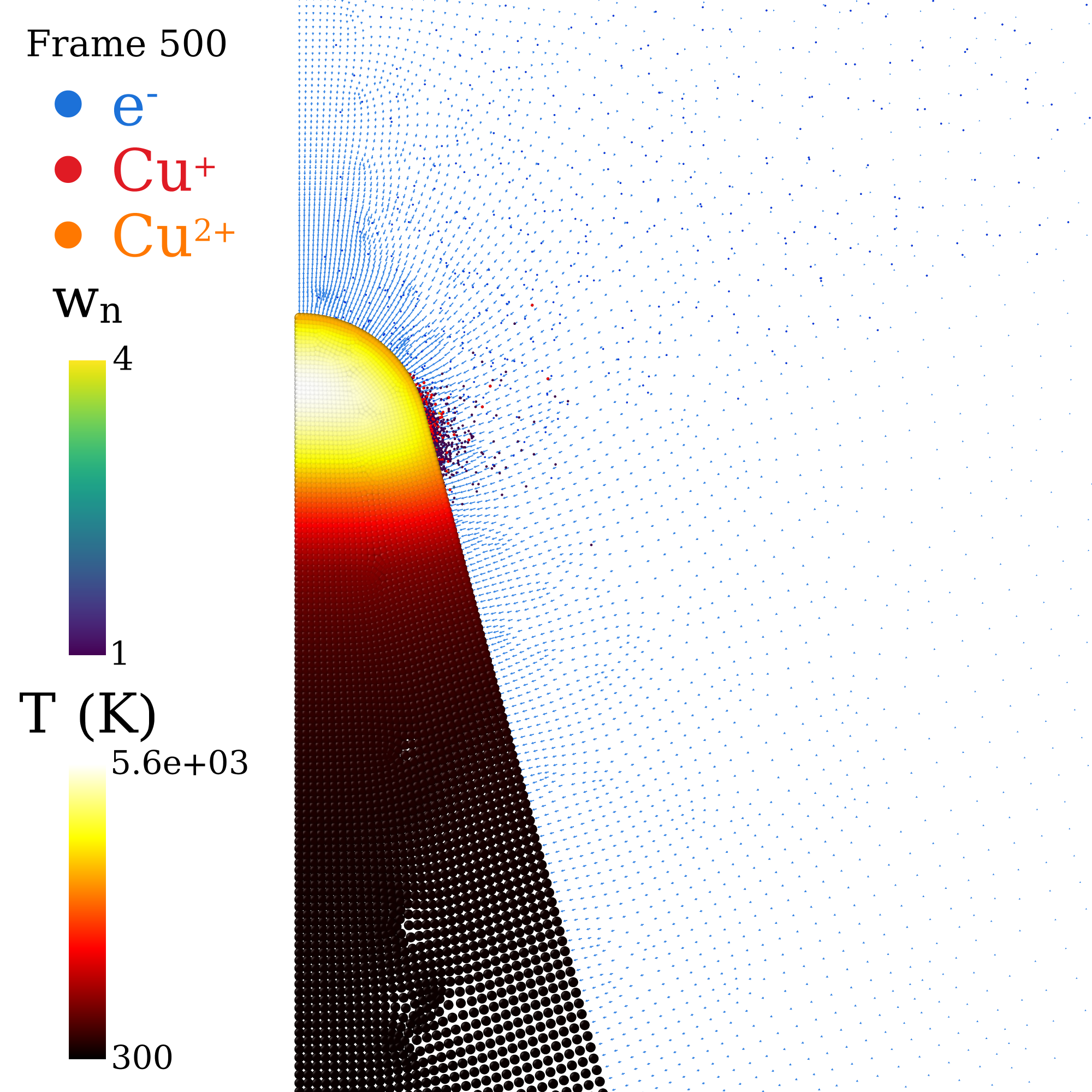}{At $t = 125 \us{ps}.$\label{fig:state15t1}}
{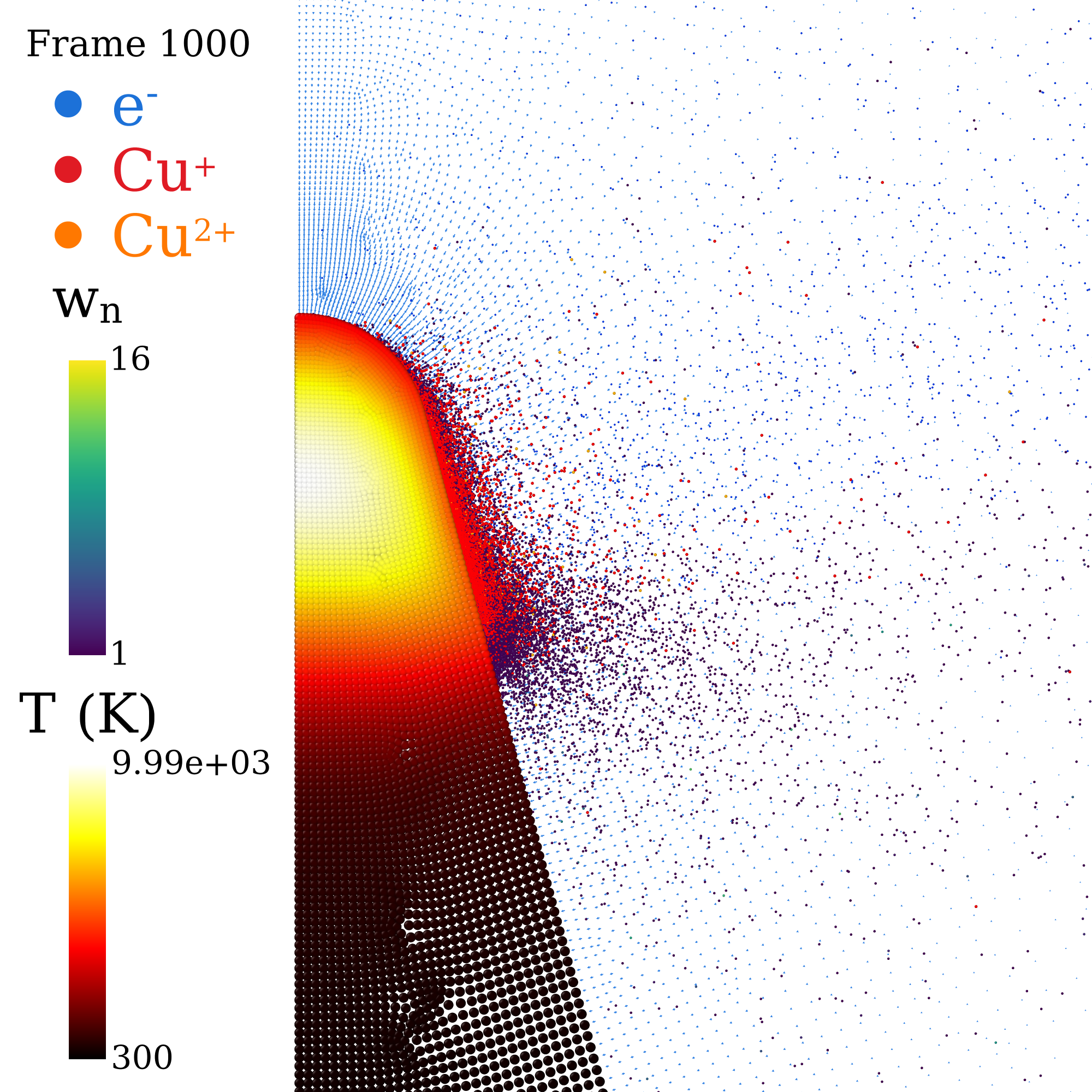}{At $t = 250 \us{ps}.$\label{fig:state15t2}}
{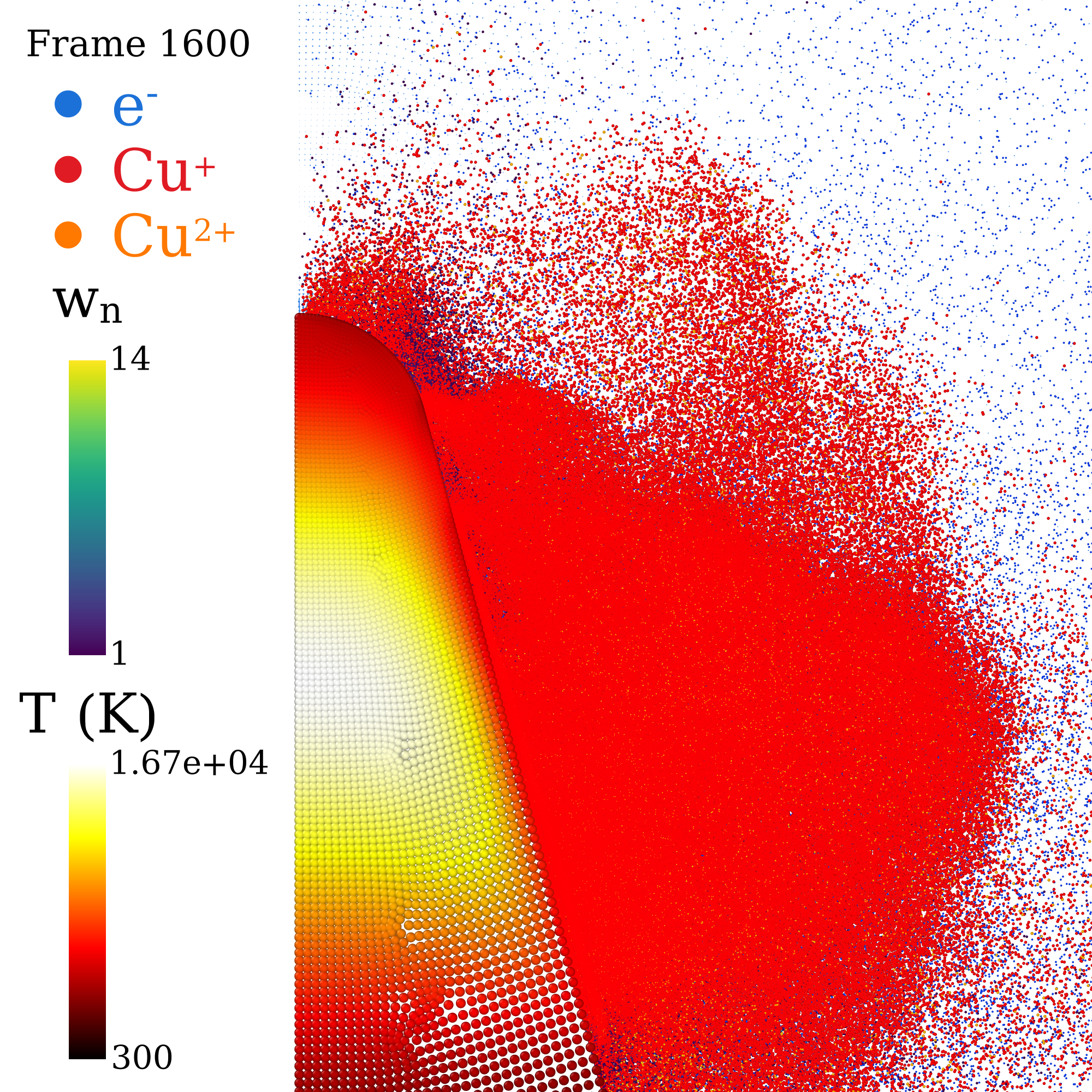}{At $t = 400 
\us{ps}.$\label{fig:state15t3}}
{Visualization of the top of the nanotip with a local field of $F_\x{loc} = 15 \us{GV/m}$. These figures combine the temperature distribution in the bulk (mesh points), electric field in the vacuum (blue arrows) and positions of SPs (spheres).}{1.0}

The state of the simulated system is shown in figures \ref{fig:state15t1}--\ref{fig:state15t3} at different times for $F_\x{loc} = 15 \us{GV/m}$. Animations for both the $F_\x{loc} = 13 \us{GV/m}$ and $F_\x{loc} = 15 \us{GV/m}$ cases are available in the supplementary material. In these figures, we show the temperature distribution in the bulk, as well as the positions of superparticles. Neutral SPs are colored according to their weights $w_n$, while electron SPs are shown in blue, $\cu^+$ ions in red and $\cu^{2+}$ ions in orange. We can see the cloud of ions expanding, with a rapid increase in temperature.

\trifighs{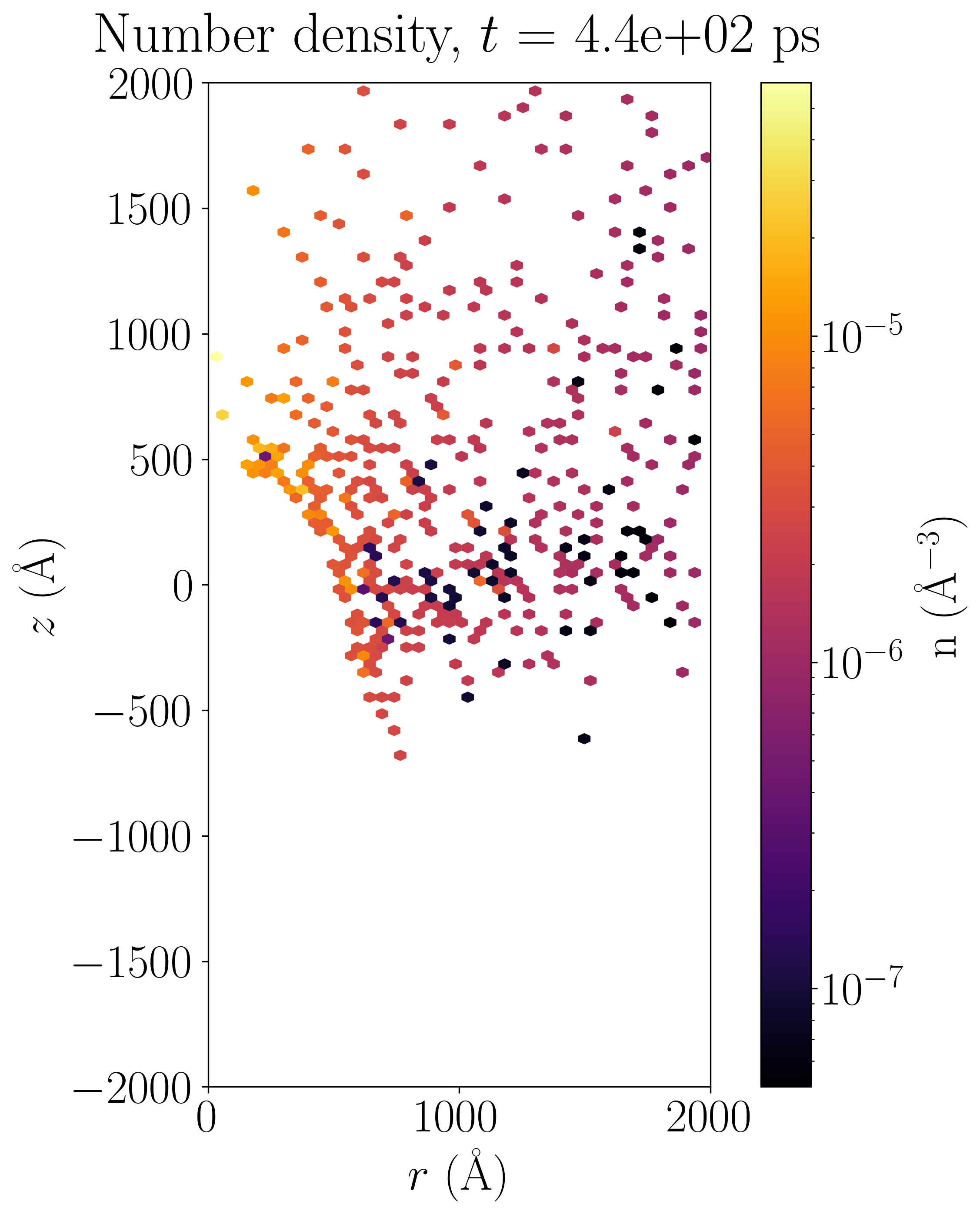}{Electrons.\label{fig:en13}}
{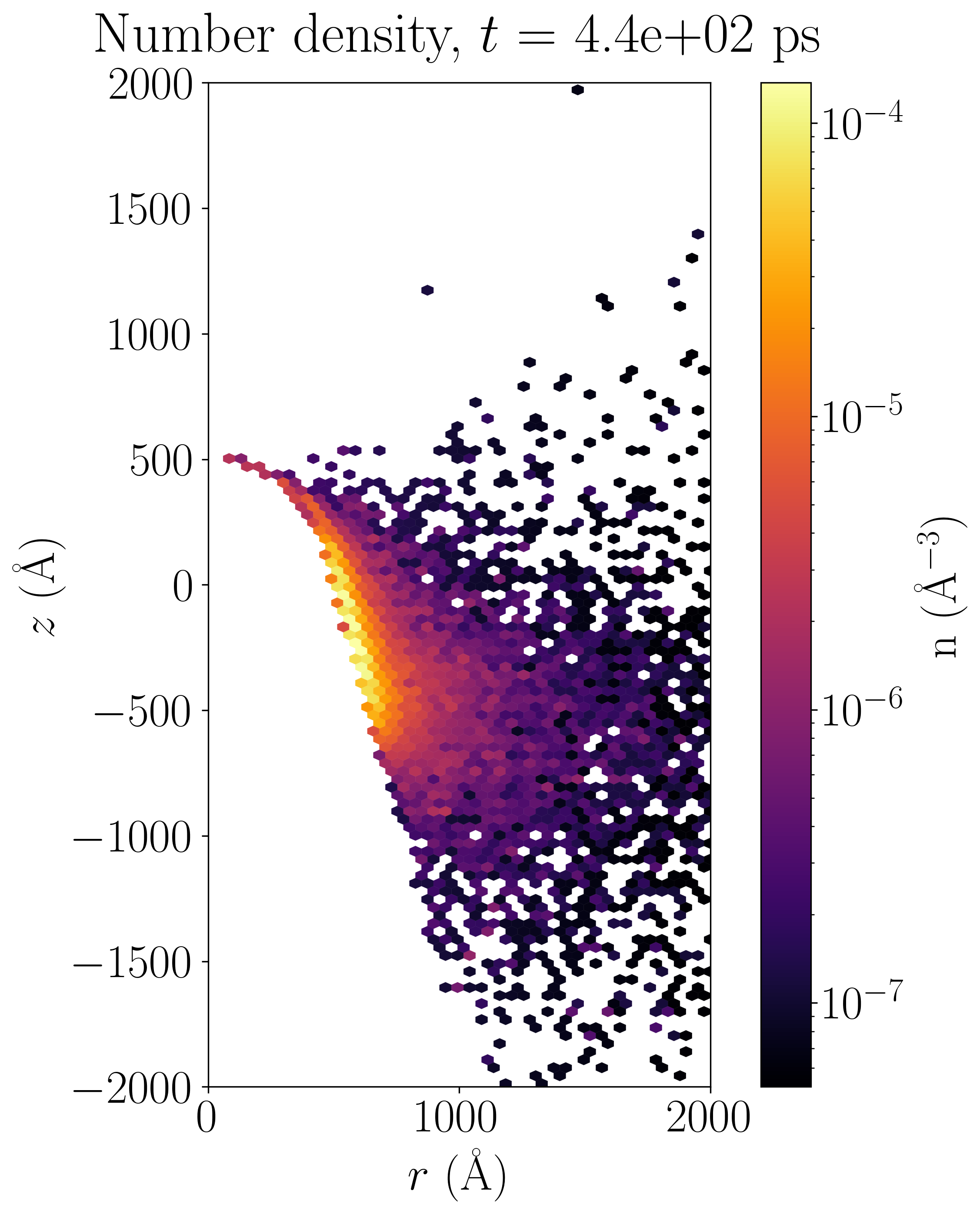}{Neutrals.\label{fig:nn13}}
{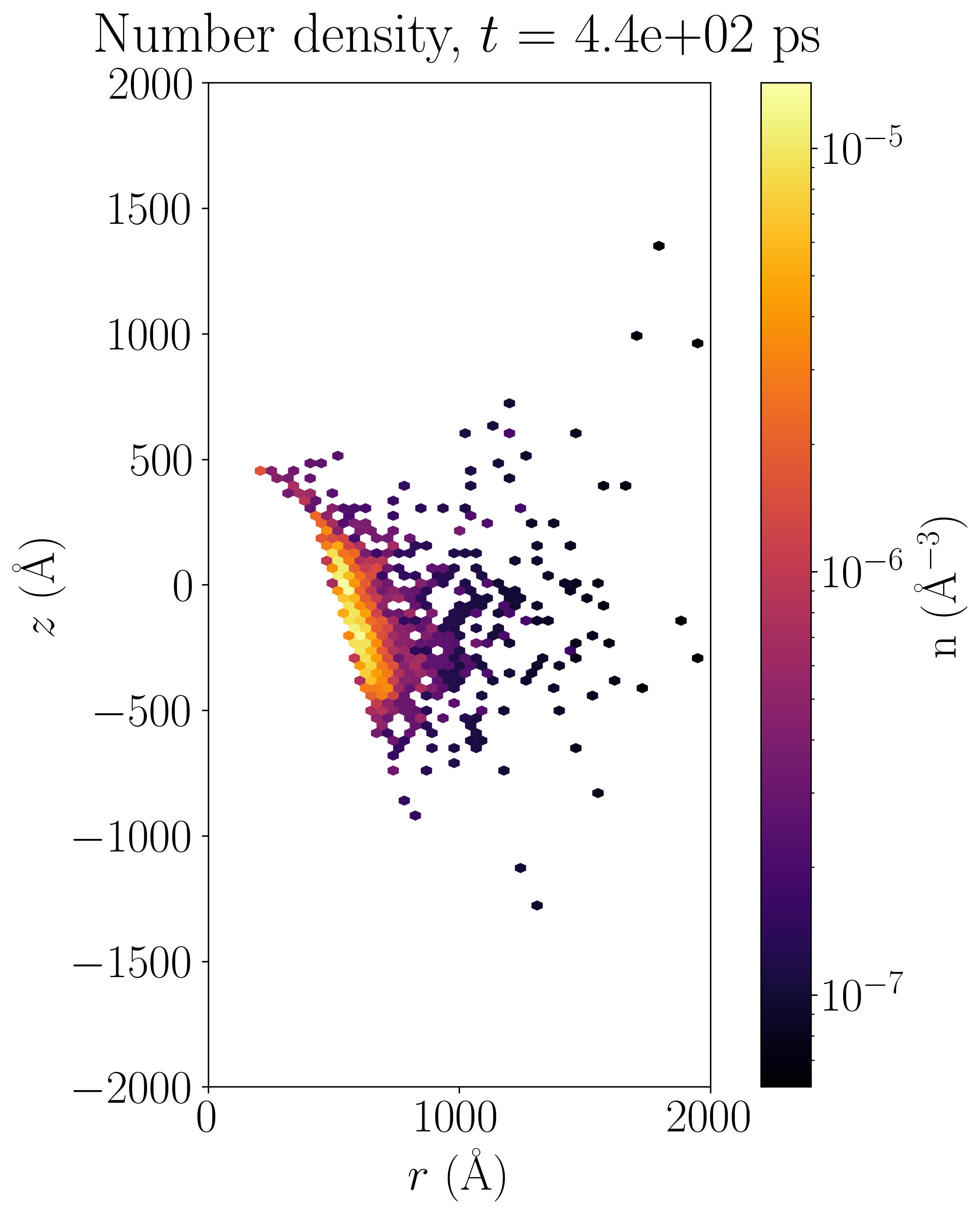}{Cu$^+$ ions.\label{fig:in13}}
{Number density distributions for different particles at a local field of $F_\x{loc} = 13 \us{GV/m}$.}{1.1}
\trifighs{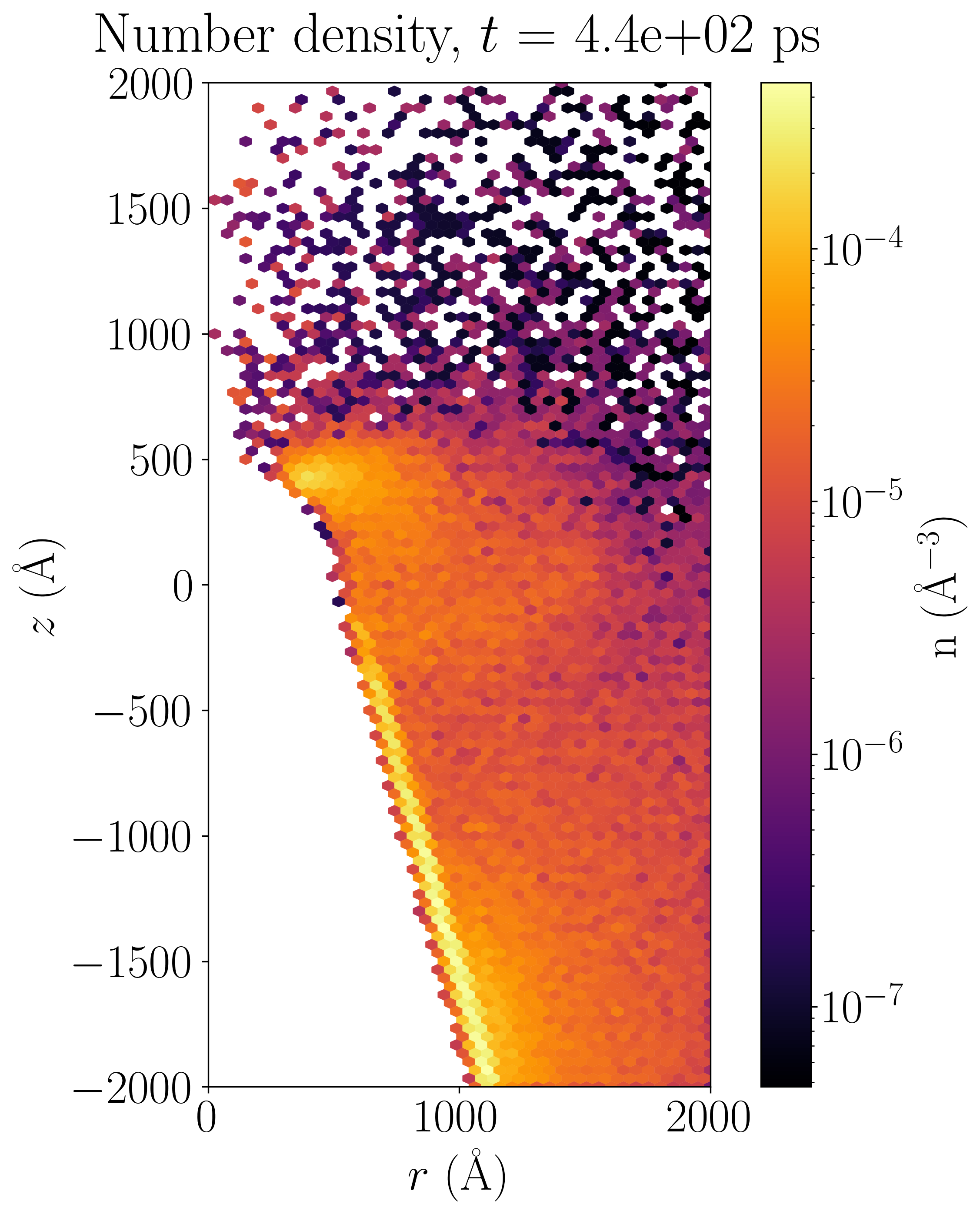}{Electrons.\label{fig:en15}}
{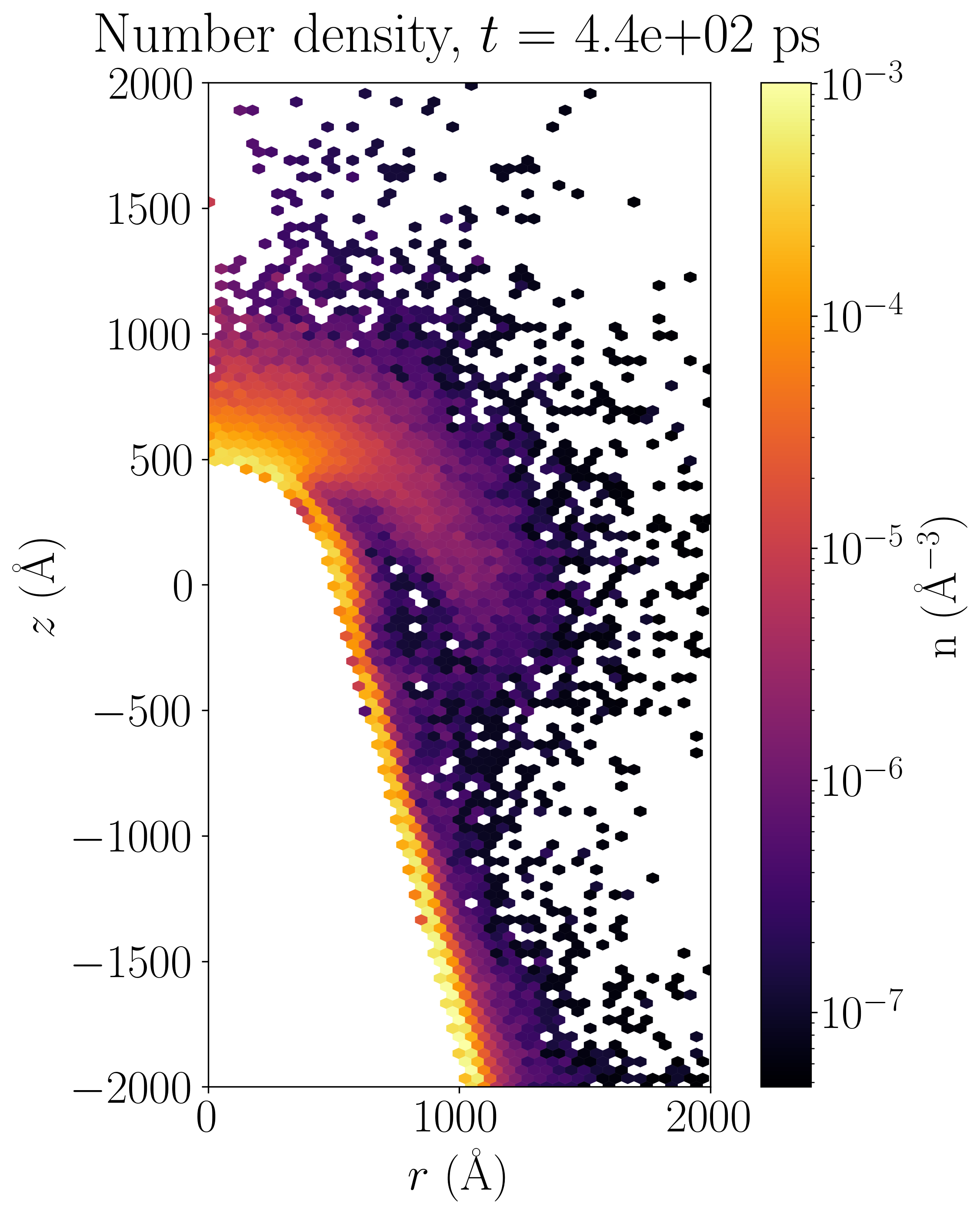}{Neutrals.\label{fig:nn15}}
{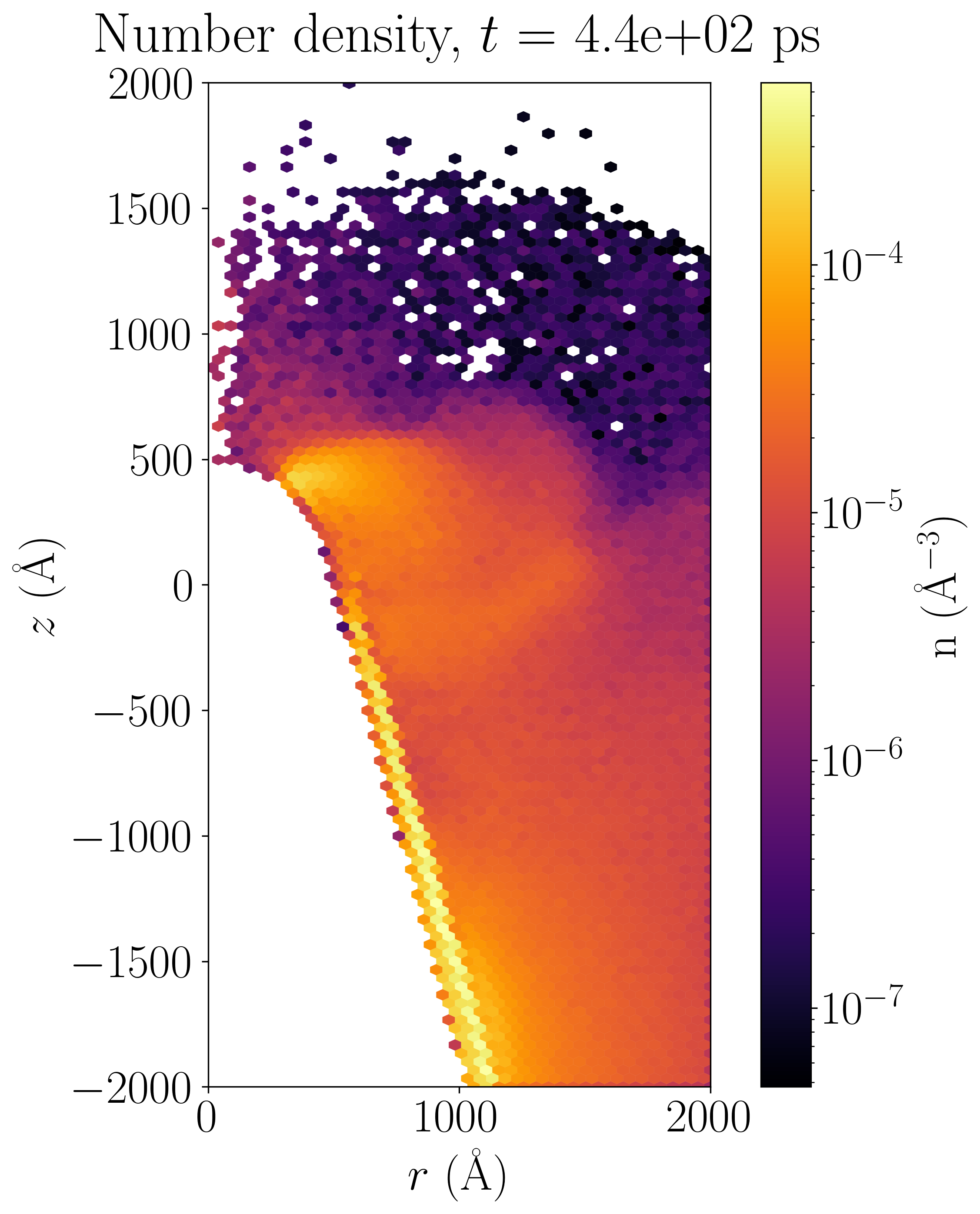}{Cu$^+$ ions.\label{fig:in15}}
{Number density distributions for different particles at a local field of $F_\x{loc} = 15 \us{GV/m}$.}{1.1}

\trifighs{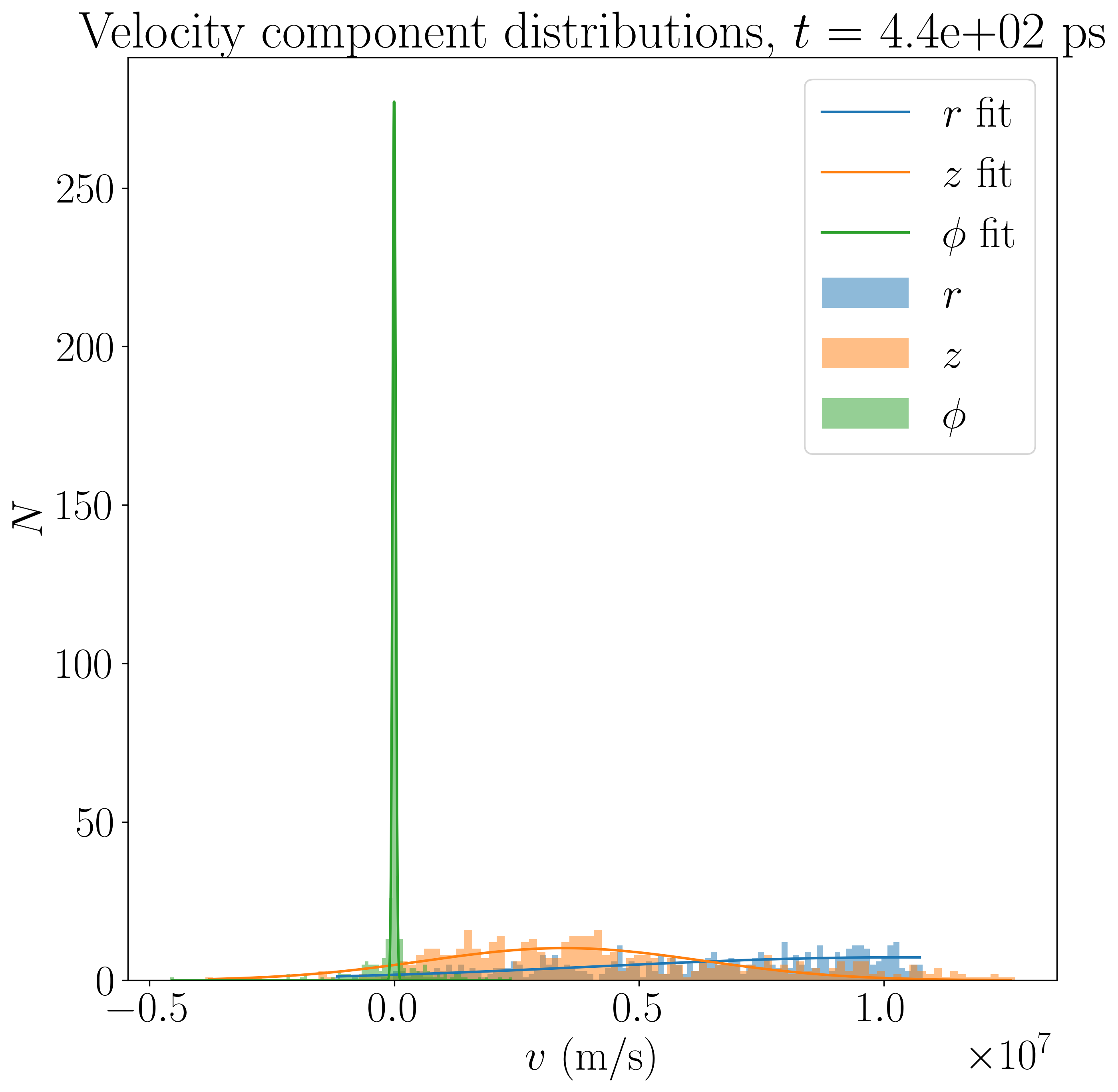}{Electrons.\label{fig:ev13}}
{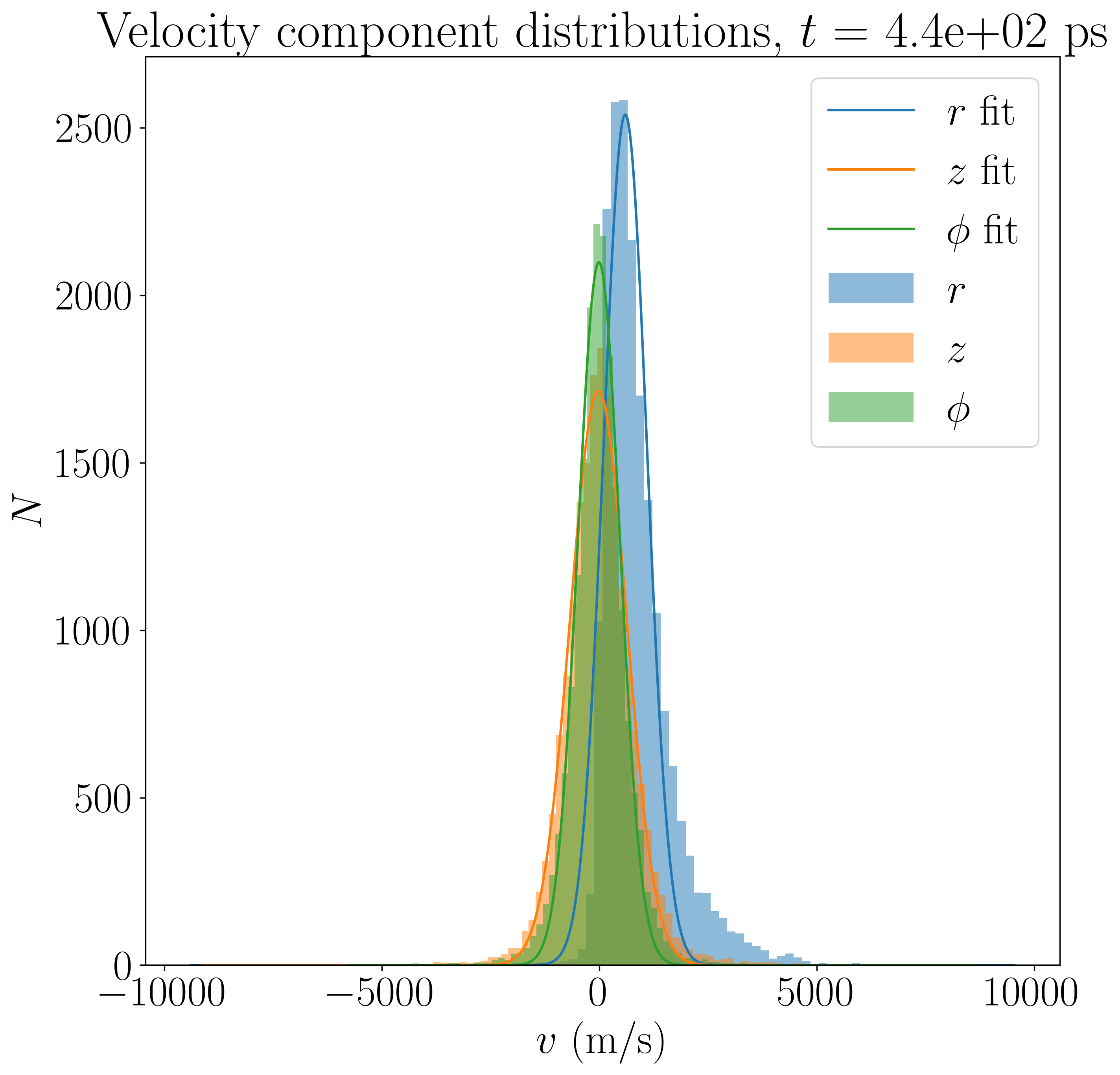}{Neutrals.\label{fig:nv13}}
{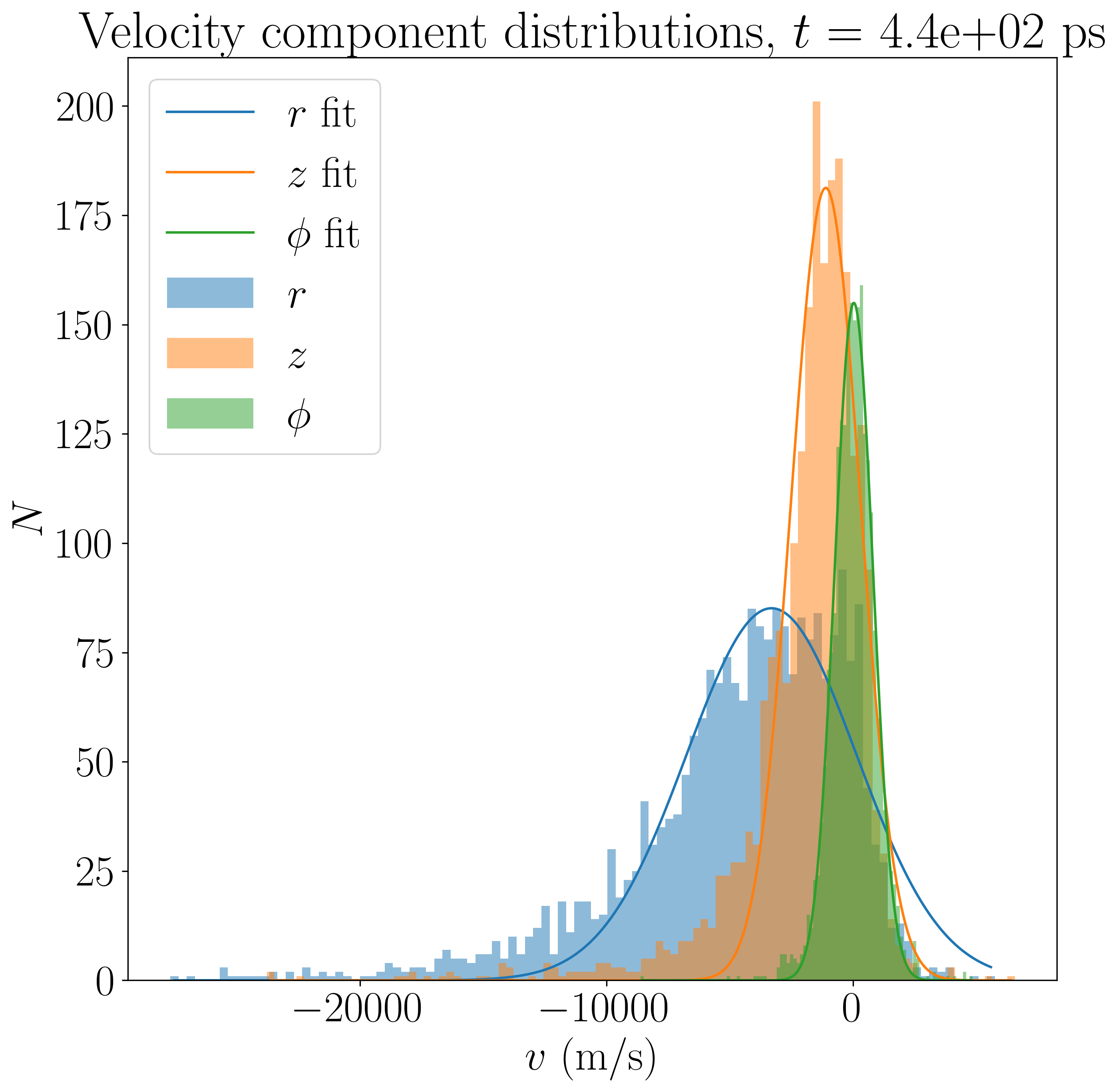}{Cu$^+$ ions.\label{fig:iv13}}
{Velocity distributions for different particles at a local field of $F_\x{loc} = 13 \us{GV/m}$.}{0.95} 
\trifighs{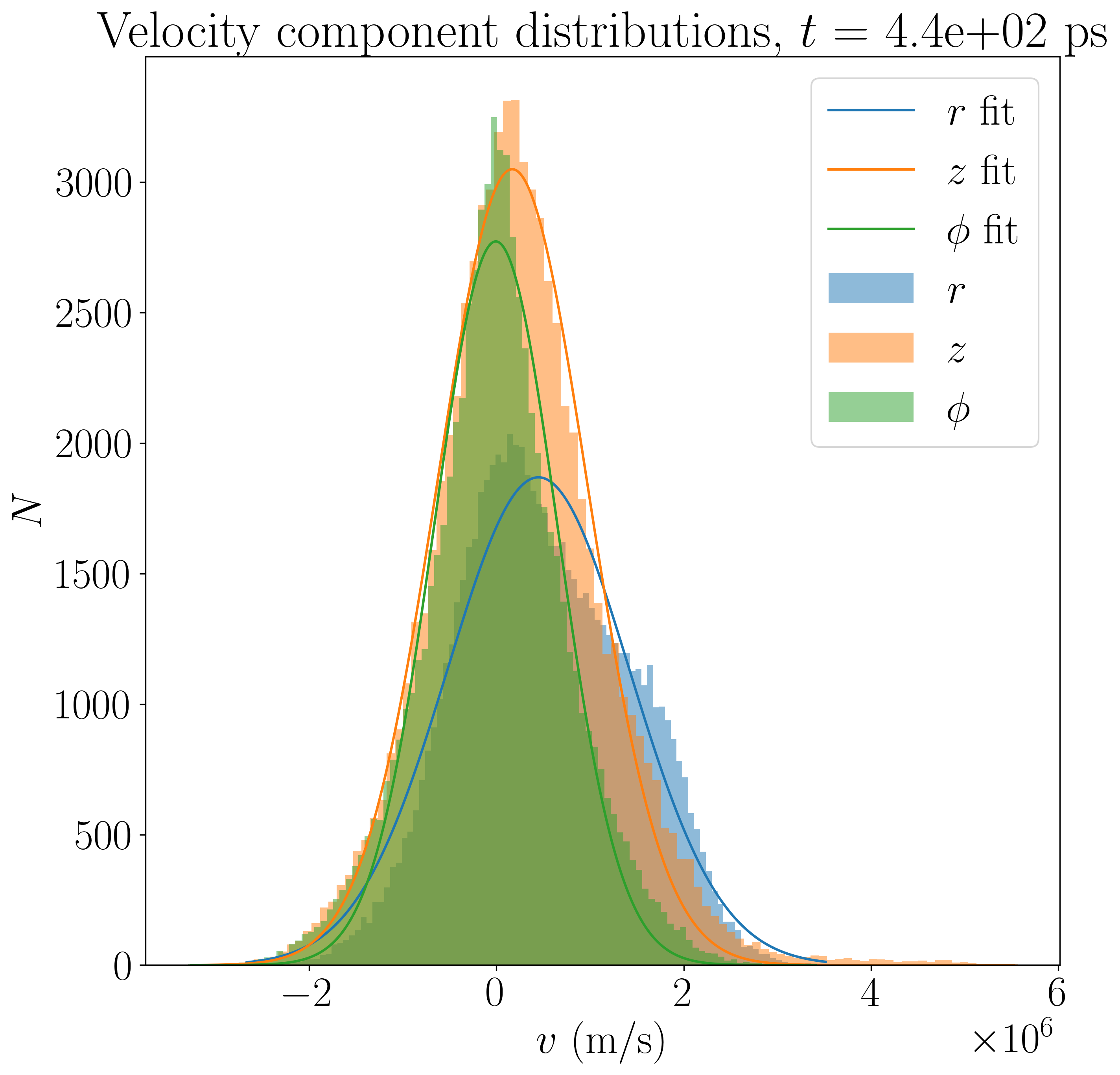}{Electrons.\label{fig:ev15}}
{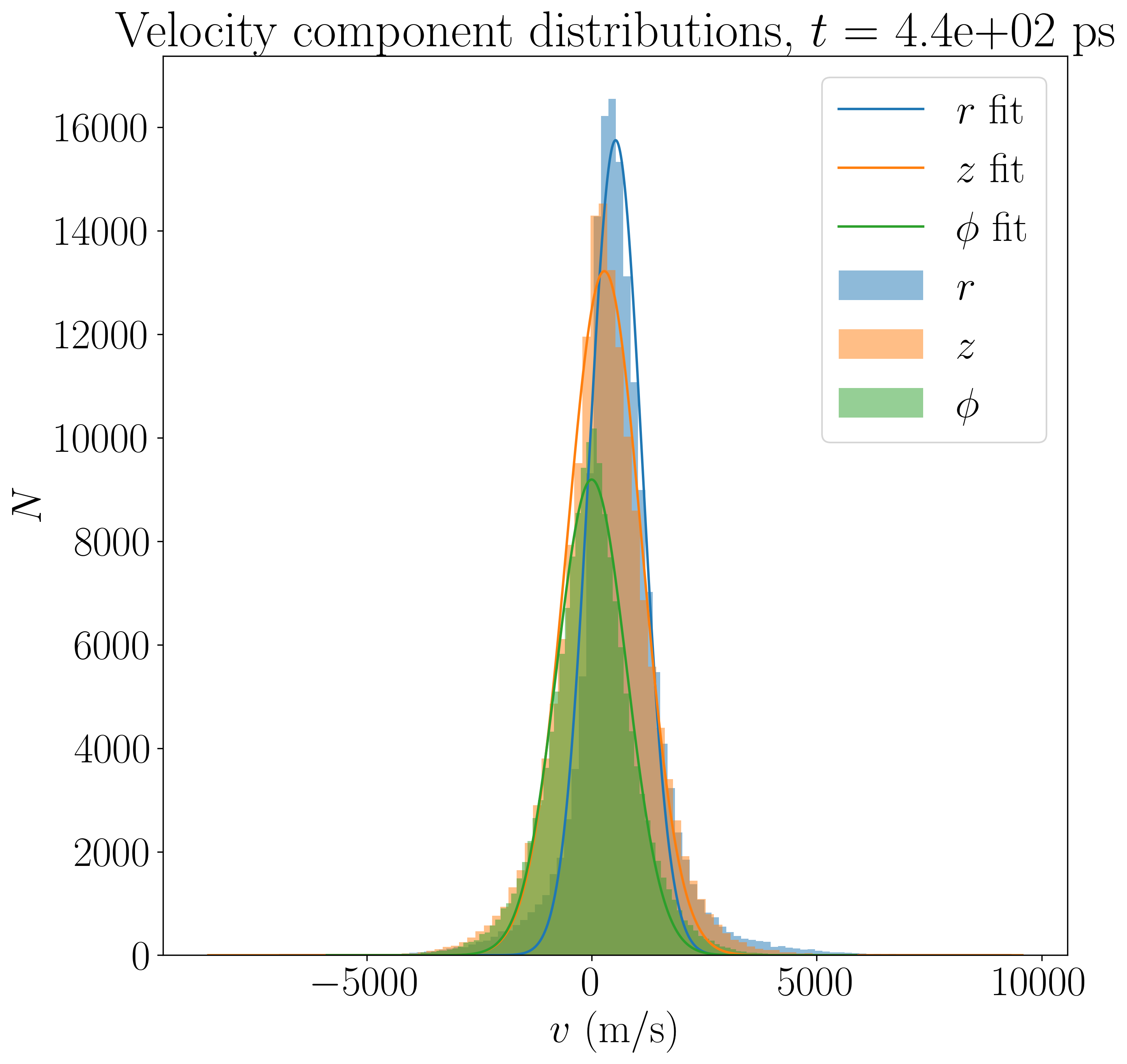}{Neutrals.\label{fig:nv15}}
{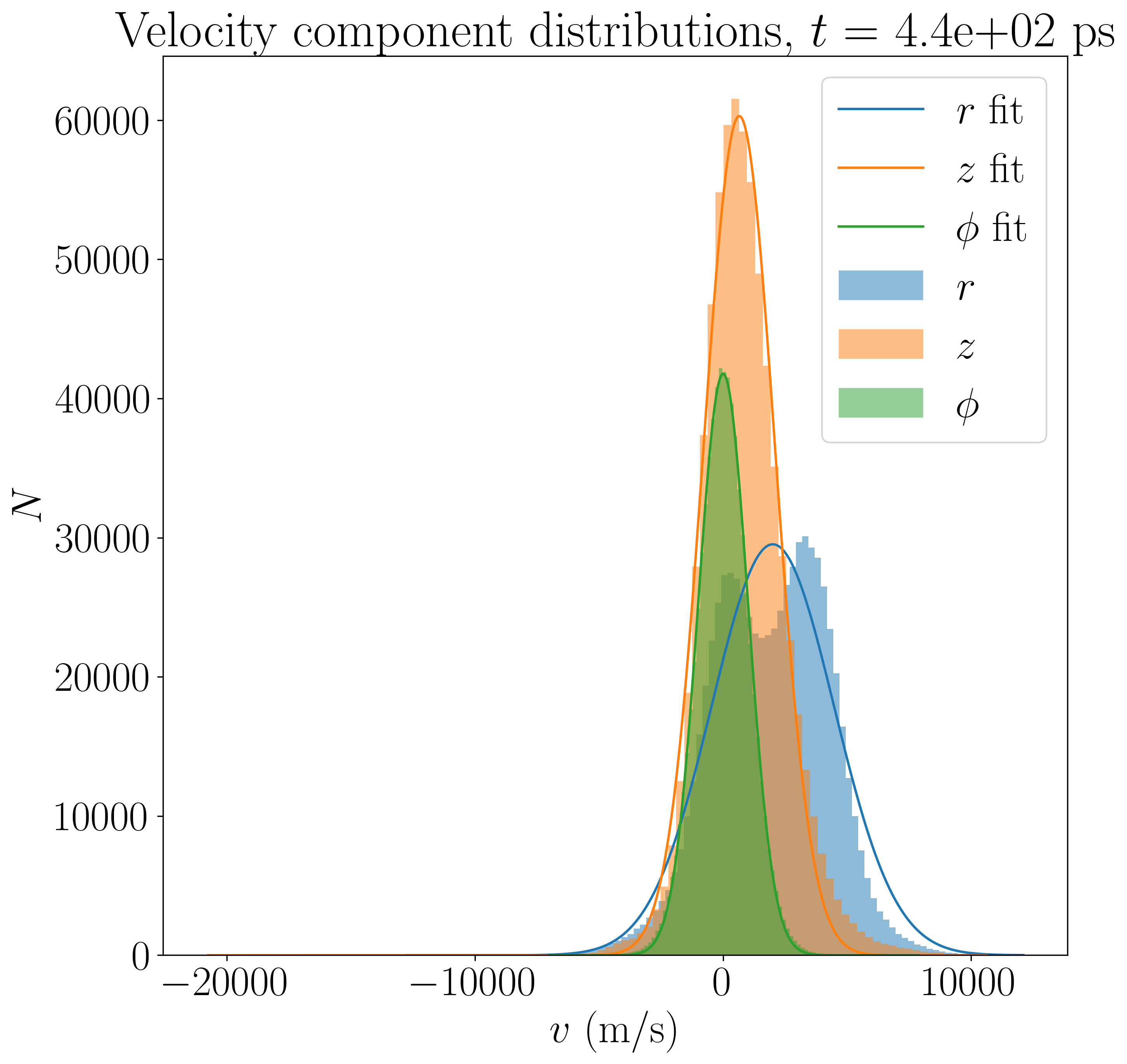}{Cu$^+$ ions.\label{fig:iv15}}
{Velocity distributions for different particles at a local field of $F_\x{loc} = 15 \us{GV/m}$.}{0.95}

In figures \ref{fig:en13}--\ref{fig:in15}, we have plotted the number density distributions for different particle species. These are the distributions for actual particles, so SP weights are accounted for. Because the geometry is axisymmetric, the factor $2\pi r$ is taken into account when calculating the number density. We can see that at a lower field a large number of neutrals is present, with only a few ions being produced. At a larger field, the number of electrons and ions is much higher. Ions (and electrons) form from neutrals by ionization, so at a higher temperature (with more evaporation) there are more neutrals available above the cathode surface. These ions are accelerated towards the cathode and bombard the surface, so before plasma formation the number of ions present remains low. In figure \ref{fig:nn13}, we can see that the number of neutrals at the top of the cathode is low, which is explained by the temperature distribution as well as the formation of ions. Since the field is highest at the top and there are many emitted electrons, neutrals are most likely ionized early on both by electron impacts and by the field directly. In figure \ref{fig:nn15}, we can see that at a larger field, the number density of neutrals increases faster. In the distribution, there is a gap corresponding to the ion distribution in figure \ref{fig:in15}. The electron and ion distributions have roughly the same shape, which suggests that the electrons in the plasma were produced by ionizations. The number densities are close to identical, which means a total charge close to 0, indicating a neutral plasma.

In figures \ref{fig:ev13}--\ref{fig:iv15}, the velocity component distributions are plotted for different particle species. For particle speeds following the Maxwell-Boltzmann distribution, the velocity components should be Gaussian. For each of the distributions, the best Gaussian fit is also shown. In figure \ref{fig:ev13}, we can see that the electron distribution before plasma formation is shifted to the right, which shows that emitted electrons are moving away from the cathode, accelerated by the field. The $\phi$-component is close to 0 for almost all electrons because almost all electrons are produced by emission, not by collisions. Neutrals in figure \ref{fig:nv13} show Gaussian distributions centered near 0, with the $r$-component distribution resulting from the simulation geometry. The ion distributions in figure \ref{fig:iv13} are shifted to the left, meaning ions are accelerated towards the cathode by the field, bombarding the cathode surface. The mean speed of the ions (accelerated towards the cathode) for 13~GV/m is about 5000~m/s (from the fitted Gaussian). When we increase the field so that plasma starts forming, these distributions change. In figure \ref{fig:ev15} the electrons are no longer moving outwards like before, which suggests they most likely originate from ionizations. These electrons are present in the cloud of plasma alongside ions, as can be seen in figure \ref{fig:en15}. In figure \ref{fig:iv15} we can see that the ion velocity distributions are shifted to the right, which shows that ions are moving away from the cathode. This can also be observed in figures \ref{fig:state15t1}--\ref{fig:state15t3}. Additionally, we note that the $r$-component distribution in figure \ref{fig:iv15} has a second peak, which suggests the influence of ion interactions. The outer layers of the plasma cloud are likely accelerated outwards in the radial direction by the large number of ions close to the cathode surface. The mean speed of the ions (expanding outwards) for 15~GV/m is about 2000~m/s (from the fitted Gaussian). The second peak shows that the plasma is expanding at a speed of slightly less than 5000~m/s. In experimental measurements, the expansion velocity in breakdown has been measured for Al foil at well above 10~km/s, so this process is likely at an early stage \cite{rouexp}.

\section{Discussion}\label{sec:discussion}

The methodological development we have described is a significant step towards a simulation model that combines emission, heating and plasma simulation. We have introduced interactions that have been neglected in previous simulations, but evidently have a significant impact on the vacuum arcing process, and especially the plasma initiation.

In our static nanotip simulations, we found that an external local field approximately in the range of $10-15 \us{GV/m}$ is sufficient to cause thermal runaway to occur. For our simulation with a local field of $10 \us{GV/m}$, plasma formation was not observed within $1 \us{ns}$. For local fields of $13 \us{GV/m}$ and $15 \us{GV/m}$, plasma initiation is observed within $1 \us{ns}$. At these external fields, the local surface field after electron emission starts is approximately $10 \us{GV/m}$, as seen in figures \ref{fig:tjf13} and \ref{fig:tjf15}. On longer timescales, a lower field may be sufficient. Additionally, the available energy for the plasma depends on the external field. For larger fields, particles in the plasma experience more acceleration, which can cause more energetic particle interactions and bombardment. Furthermore, the shape of the nanotip plays an important role on its heating dynamics (see figures \ref{fig:tdist5_5ns} and \ref{fig:tdist25_5ns}), which would also impact the exact field value needed to initiate plasma. In previous simulations it has been observed that at high temperatures nanotips can sharpen, which increases the local field \cite{femocs}. In experiments, a local field of $10.8 \us{GV/m}$ has been observed to cause DC vacuum arcs \cite{desarc}. In general, our simulations show that an extremely high temperature on the order of several thousands of degrees is needed to produce the evaporation rate that is necessary for the initiation of plasma.

Since our model is static, the thermal runaway process does not change the shape of the nanotip. In our simulations, the nanotip is seen reaching temperatures above $5000 \us{K}$, at which point the nanotip would already rapidly vaporize. This means that beyond the initial stages of the runaway process, it is uncertain how the nanotip evolves and how the shape of the nanotip influences the runaway process as a whole. When we have neutral evaporation or sputtering, atoms would be ejected from the surface and the nanotip would shrink. In our current model these processes are not limited, so the number of particles in the system continues to increase while the nanotip retains its original form. In the simulation, the number of neutrals and ions in the plasma is approximately $8.2 \tp{6}$. We can calculate the number of Cu atoms in the nanotip to be about $1.1 \tp{11}$, which means that at this stage the plasma contains only a small fraction of the material in the tip. For a nanotip scaled to $1/10$ the size, the number of Cu atoms in the nanotip is $1.1 \tp{8}$, still less than in the plasma cloud at this stage. To model this process in detail, full coupling with molecular dynamics is needed. This is an especially important direction for future work towards the development of a more complete model.

A model that tracks the evolution of the tip shape during the thermal runaway process has been developed previously in \cite{kyrhea}. Deformation of the nanotip when heated can significantly influence the surface morphology and therefore the local field on the metal surface. However, this model disregarded the exact plasma formation processes. The present work provides the next step towards full integration between plasma simulations and molecular dynamics, although introducing the exchange processes between material (MD) and plasma (PIC) is the subject of ongoing and future research.

The focus of the present work has been to model the interactions that couple the plasma to the surface and pinpoint the ones that are essential for vacuum arc initiation. Our results show that ion bombardment contributes to surface heating. Sputtering, which has been previously considered a major source of neutral vapor from the surface \cite{arcpic}, was found to be much less important than evaporation during arc initiation. This corroborates the importance of the previously used ``high flux sputtering'' approximation to provide neutrals in the ArcPIC model \cite{arcpic}. Cathode surface interactions could lead to more evaporation/sputtering and more bombardment, forming a positive feedback loop that can significantly influence the development of the system. Furthermore, direct field ionization was found to play a very significant role in the plasma formation process, as it provides the very initial ionization of neutrals.

Finally, a significant factor that affects the plasma formation dynamics is the coupling of the vacuum arc initiation spot to the electromagnetic power available in a high-field device \cite{wuepow, paspow, grupow, wanpow}. In order to better model the power coupling, a circuit model taking into account impedance changes on the cathode surface would be needed.

\section{Conclusions}\label{sec:conclusions}

Our simulations show that even a static intensively field-emitting nanotip can go into thermal runaway and form vacuum arc plasma around it if a sufficiently high electric field is applied. Heating of the cathode surface results in an increasing rate of evaporating neutrals from the surface, which then ionize by electron impacts and by the field directly, leading to ionized plasma formation and a large discharge current between the cathode and the anode. Within our static simulation model, we identified the most important processes for vacuum arcing. We found that direct field ionization is more significant compared to electron impact ionization at the beginning of plasma formation, which is an aspect of plasma formation that has been overlooked in previous simulations. Furthermore, the influx of neutral vapor from the surface into the plasma was found to be dominated by thermal evaporation, while sputtering played a secondary role. Nottingham heating, evaporative cooling and bombardment heating were all found to contribute significantly to the heating dynamics of the cathode surface. In order to fully simulate the dynamic evolution of the cathode surface, molecular dynamics is needed. Adding coupling between our current plasma simulation and molecular dynamics is the main direction for future work.

\section*{Acknowledgements}

We wish to thank Mihkel Veske for useful discussions and major contributions to the FEMOCS code.

RK, AK, TT and VZ are supported by the European Union's Horizon 2020 research and innovation program, under Grant Agreement No. 856705 (ERA Chair ``MATTER'') and by the Estonian Research Council's project RVTT3 – ``CERN Science Consortium of Estonia''. TT is supported by the Estonian Research Council Grant No. SJD61. RK, AK and FD have been supported by CERN CLIC K-contract (No. 47207461). RK is supported by the doctoral program MATRENA of the University of Helsinki. Computational resources provided by Finnish Grid and Cloud Infrastructure (persistent identifier urn:nbn:fi:research-infras-2016072533).

\bibliographystyle{elsarticle-num}
\bibliography{vacuum_arc_plasma_initiation}

\begin{thebibliography}{10}
\expandafter\ifx\csname url\endcsname\relax
  \def\url#1{\texttt{#1}}\fi
\expandafter\ifx\csname urlprefix\endcsname\relax\def\urlprefix{URL }\fi
\expandafter\ifx\csname href\endcsname\relax
  \def\href#1#2{#2} \def\path#1{#1}\fi

\bibitem{clic}
{CERN}, {CERN Yellow Reports: Monographs, Vol 4 (2018): The Compact Linear
  Collider (CLIC) - Project Implementation Plan} (2019).
\newblock \href {https://doi.org/10.23731/CYRM-2018-004}
  {\path{doi:10.23731/CYRM-2018-004}}.

\bibitem{boxarc}
R.~L. Boxman, D.~M. Sanders, P.~J. Martin, {Handbook of Vacuum Arc Science and
  Technology: Fundamentals and Applications}, Noyes Publications, Park Ridge,
  NJ, 1995.

\bibitem{arcpic}
H.~Timko, K.~Ness~Sjobak, L.~Mether, S.~Calatroni, F.~Djurabekova, K.~Matyash,
  K.~Nordlund, R.~Schneider, W.~Wuensch, From field emission to vacuum arc
  ignition: A new tool for simulating copper vacuum arcs, Contributions to
  Plasma Physics 55~(4) (2015) 299--314.
\newblock \href {https://doi.org/10.1002/ctpp.201400069}
  {\path{doi:10.1002/ctpp.201400069}}.

\bibitem{femocs}
M.~Veske, A.~Kyritsakis, F.~Djurabekova, K.~N. Sjobak, A.~Aabloo, V.~Zadin,
  Dynamic coupling between particle-in-cell and atomistic simulations, Phys.
  Rev. E 101 (2020) 053307.
\newblock \href {https://doi.org/10.1103/PhysRevE.101.053307}
  {\path{doi:10.1103/PhysRevE.101.053307}}.

\bibitem{zhoexp}
Z.~Zhou, A.~Kyritsakis, Z.~Wang, Y.~Li, Y.~Geng, F.~Djurabekova, Direct
  observation of vacuum arc evolution with nanosecond resolution, Scientific
  Reports 9~(1) (2019) 7814.
\newblock \href {https://doi.org/10.1038/s41598-019-44191-6}
  {\path{doi:10.1038/s41598-019-44191-6}}.

\bibitem{jandif}
V.~Jansson, E.~Baibuz, A.~Kyritsakis, S.~Vigonski, V.~Zadin, S.~Parviainen,
  A.~Aabloo, F.~Djurabekova, Growth mechanism for nanotips in high electric
  fields, Nanotechnology 31~(35) (2020) 355301.
\newblock \href {https://doi.org/10.1088/1361-6528/ab9327}
  {\path{doi:10.1088/1361-6528/ab9327}}.

\bibitem{kimdif}
J.~Kimari, Y.~Wang, A.~Kyritsakis, V.~Zadin, F.~Djurabekova, Biased
  self-diffusion on cu surface due to electric field gradients, Journal of
  Physics D: Applied Physics 55~(46) (2022) 465302.
\newblock \href {https://doi.org/10.1088/1361-6463/ac91dd}
  {\path{doi:10.1088/1361-6463/ac91dd}}.

\bibitem{pohdlo}
A.~S. Pohjonen, S.~Parviainen, T.~Muranaka, F.~Djurabekova, Dislocation
  nucleation on a near surface void leading to surface protrusion growth under
  an external electric field, Journal of Applied Physics 114~(3) (2013) 033519.
\newblock \href {https://doi.org/10.1063/1.4815938}
  {\path{doi:10.1063/1.4815938}}.

\bibitem{zaddlo}
V.~Zadin, A.~Pohjonen, A.~Aabloo, K.~Nordlund, F.~Djurabekova,
  Electrostatic-elastoplastic simulations of copper surface under high electric
  fields, Phys. Rev. ST Accel. Beams 17 (2014) 103501.
\newblock \href {https://doi.org/10.1103/PhysRevSTAB.17.103501}
  {\path{doi:10.1103/PhysRevSTAB.17.103501}}.

\bibitem{sarexp}
A.~Saressalo, A.~Kyritsakis, F.~Djurabekova, I.~Profatilova, J.~Paszkiewicz,
  S.~Calatroni, W.~Wuensch, Classification of vacuum arc breakdowns in a pulsed
  dc system, Phys. Rev. Accel. Beams 23 (2020) 023101.
\newblock \href {https://doi.org/10.1103/PhysRevAccelBeams.23.023101}
  {\path{doi:10.1103/PhysRevAccelBeams.23.023101}}.

\bibitem{wuepow}
W.~Wuensch, \href{https://cds.cern.ch/record/932674}{{The Scaling of the
  Traveling-Wave RF Breakdown Limit}}, Tech. rep., CERN, Geneva (2006).
\newline\urlprefix\url{https://cds.cern.ch/record/932674}

\bibitem{paspow}
J.~Paszkiewicz, A.~Grudiev, W.~Wuensch, Local power coupling as a predictor of
  high-gradient breakdown performance (2022).
\newblock \href {http://arxiv.org/abs/2209.15291} {\path{arXiv:2209.15291}}.

\bibitem{grupow}
A.~Grudiev, S.~Calatroni, W.~Wuensch, New local field quantity describing the
  high gradient limit of accelerating structures, Phys. Rev. ST Accel. Beams 12
  (2009) 102001.
\newblock \href {https://doi.org/10.1103/PhysRevSTAB.12.102001}
  {\path{doi:10.1103/PhysRevSTAB.12.102001}}.

\bibitem{wanpow}
D.~Wang, A.~Kyritsakis, A.~Saressalo, L.~Wang, F.~Djurabekova, {Effects of the
  electromagnetic power coupling on vacuum breakdown}, Vacuum 210 (Apr 2023).
\newblock \href {https://doi.org/10.1016/j.vacuum.2023.111880}
  {\path{doi:10.1016/j.vacuum.2023.111880}}.

\bibitem{desarc}
A.~Descoeudres, Y.~Levinsen, S.~Calatroni, M.~Taborelli, W.~Wuensch,
  Investigation of the {DC} vacuum breakdown mechanism, Phys. Rev. ST Accel.
  Beams 12 (2009) 092001.
\newblock \href {https://doi.org/10.1103/PhysRevSTAB.12.092001}
  {\path{doi:10.1103/PhysRevSTAB.12.092001}}.

\bibitem{wiefra}
R.~Franz, G.~Wiedemann, {Ueber die W\"arme-Leitungsf\"ahigkeit der Metalle},
  Annalen der Physik 165~(8) (1853) 497--531.
\newblock \href {https://doi.org/10.1002/andp.18531650802}
  {\path{doi:10.1002/andp.18531650802}}.

\bibitem{natlon}
P.~Nath, K.~Chopra, Thermal conductivity of copper films, Thin Solid Films
  20~(1) (1974) 53--62.
\newblock \href {https://doi.org/10.1016/0040-6090(74)90033-9}
  {\path{doi:10.1016/0040-6090(74)90033-9}}.

\bibitem{gmsh}
C.~Geuzaine, J.-F. Remacle, {Gmsh: A 3-D finite element mesh generator with
  built-in pre- and post-processing facilities}, International Journal for
  Numerical Methods in Engineering 79~(11) (2009) 1309--1331.
\newblock \href {https://doi.org/10.1002/nme.2579}
  {\path{doi:10.1002/nme.2579}}.

\bibitem{dealii}
D.~Arndt, W.~Bangerth, B.~Blais, M.~Fehling, R.~Gassm{\"o}ller, T.~Heister,
  L.~Heltai, U.~K{\"o}cher, M.~Kronbichler, M.~Maier, P.~Munch, J.-P. Pelteret,
  S.~Proell, K.~Simon, B.~Turcksin, D.~Wells, J.~Zhang, The \texttt{deal.II}
  library, version 9.3, Journal of Numerical Mathematics 29~(3) (2021)
  171--186.
\newblock \href {https://doi.org/10.1515/jnma-2021-0081}
  {\path{doi:10.1515/jnma-2021-0081}}.

\bibitem{birpic}
C.~K. Birdsall, A.~B. Langdon, {Plasma Physics via Computer Simulation},
  McGraw-Hill, New York, NY, 1985.

\bibitem{pic}
D.~Tskhakaya, K.~Matyash, R.~Schneider, F.~Taccogna, The particle-in-cell
  method, Contributions to Plasma Physics 47~(8-9) (2007) 563--594.
\newblock \href {https://doi.org/10.1002/ctpp.200710072}
  {\path{doi:10.1002/ctpp.200710072}}.

\bibitem{apman}
K.~N. Sjobak, H.~Timko,
  \href{https://cds.cern.ch/record/1347241/files/EuCARD-NOT-2011-004.pdf}{{2D
  ArcPIC} code description: Description of methods and user/developer manual
  (second edition)}, {CERN}, 2014.
\newline\urlprefix\url{https://cds.cern.ch/record/1347241/files/EuCARD-NOT-2011-004.pdf}

\bibitem{takcol}
T.~Takizuka, H.~Abe, A binary collision model for plasma simulation with a
  particle code, Journal of Computational Physics 25~(3) (1977) 205--219.
\newblock \href {https://doi.org/10.1016/0021-9991(77)90099-7}
  {\path{doi:10.1016/0021-9991(77)90099-7}}.

\bibitem{mccpic}
V.~Vahedi, M.~Surendra, {A Monte Carlo collision model for the particle-in-cell
  method: applications to argon and oxygen discharges}, Computer Physics
  Communications 87~(1-2) (1995) 179--198.
\newblock \href {https://doi.org/10.1016/0010-4655(94)00171-W}
  {\path{doi:10.1016/0010-4655(94)00171-W}}.

\bibitem{matcol}
K.~Matyash, Kinetic modeling of multi-component edge plasmas, Ph.D. thesis,
  University of Greifswald (2003).

\bibitem{koimsc}
R.~Koitermaa, \href{https://helda.helsinki.fi/handle/10138/345533}{Concurrent
  multi-scale modelling of vacuum arc plasma initiation}, Master's thesis,
  University of Helsinki (2022).
\newline\urlprefix\url{https://helda.helsinki.fi/handle/10138/345533}

\bibitem{jengtf}
K.~L. Jensen, M.~Cahay, General thermal-field emission equation, {Applied
  Physics Letters} 88~(15) (2006) 154105.
\newblock \href {https://doi.org/10.1063/1.2193776}
  {\path{doi:10.1063/1.2193776}}.

\bibitem{eimgtf}
K.~Eimre, S.~Parviainen, A.~Aabloo, F.~Djurabekova, V.~Zadin, {Application of
  the general thermal field model to simulate the behaviour of nanoscale Cu
  field emitters}, Journal of Applied Physics 118~(3), 033303 (07 2015).
\newblock \href {https://doi.org/10.1063/1.4926490}
  {\path{doi:10.1063/1.4926490}}.

\bibitem{getelec}
A.~Kyritsakis, F.~Djurabekova, {A general computational method for electron
  emission and thermal effects in field emitting nanotips}, Computational
  Materials Science 128 (2017) 15--21.
\newblock \href {https://doi.org/10.1016/j.commatsci.2016.11.010}
  {\path{doi:10.1016/j.commatsci.2016.11.010}}.

\bibitem{zhoeva}
Z.~Zhou, A.~Kyritsakis, Z.~Wang, Y.~Li, Y.~Geng, F.~Djurabekova, Effect of the
  anode material on the evolution of the vacuum breakdown process, Journal of
  Physics D: Applied Physics 54~(3) (2020) 035201.
\newblock \href {https://doi.org/10.1088/1361-6463/abbbb7}
  {\path{doi:10.1088/1361-6463/abbbb7}}.

\bibitem{zhosr}
Z.~He, Review of the {Shockley-Ramo} theorem and its application in
  semiconductor gamma-ray detectors, Nuclear Instruments and Methods in Physics
  Research Section A: Accelerators, Spectrometers, Detectors and Associated
  Equipment 463~(1) (2001) 250--267.
\newblock \href {https://doi.org/10.1016/S0168-9002(01)00223-6}
  {\path{doi:10.1016/S0168-9002(01)00223-6}}.

\bibitem{tracs}
S.~Trajmar, W.~Williams, S.~K. Srivastava, Electron-impact cross sections for
  {Cu} atoms, Journal of Physics B: Atomic and Molecular Physics 10~(16) (1977)
  3323--3333.
\newblock \href {https://doi.org/10.1088/0022-3700/10/16/025}
  {\path{doi:10.1088/0022-3700/10/16/025}}.

\bibitem{bolcs}
M.~A. Bolorizadeh, C.~J. Patton, M.~B. Shah, H.~B. Gilbody, Multiple ionization
  of copper by electron impact, Journal of Physics B: Atomic, Molecular and
  Optical Physics 27~(1) (1994) 175--183.
\newblock \href {https://doi.org/10.1088/0953-4075/27/1/019}
  {\path{doi:10.1088/0953-4075/27/1/019}}.

\bibitem{frecs}
R.~S. Freund, R.~C. Wetzel, R.~J. Shul, T.~R. Hayes, Cross-section measurements
  for electron-impact ionization of atoms, Phys. Rev. A 41 (1990) 3575--3595.
\newblock \href {https://doi.org/10.1103/PhysRevA.41.3575}
  {\path{doi:10.1103/PhysRevA.41.3575}}.

\bibitem{pincs}
M.~S. Pindzola, D.~C. Griffin, C.~Bottcher, D.~C. Gregory, A.~M. Howald, R.~A.
  Phaneuf, D.~H. Crandall, G.~H. Dunn, D.~W. Mueller, T.~J. Morgan,
  \href{https://www.osti.gov/servlets/purl/6089462}{Survey of experimental and
  theoretical electron-impact ionization cross sections for transition metal
  ions in low stages of ionization} (1985).
\newline\urlprefix\url{https://www.osti.gov/servlets/purl/6089462}

\bibitem{grecs}
D.~C. Gregory, A.~M. Howald, Electron-impact ionization of multicharged metal
  ions: {Ni}$^{3+}$, {Cu}$^{2+}$, {Cu}$^{3+}$, and {Sb}$^{3+}$, Phys. Rev. A 34
  (1986) 97--102.
\newblock \href {https://doi.org/10.1103/PhysRevA.34.97}
  {\path{doi:10.1103/PhysRevA.34.97}}.

\bibitem{aubcs}
A.~Aubreton, M.-F. Elchinger, Transport properties in non-equilibrium argon,
  copper and argon{\textendash}copper thermal plasmas, Journal of Physics D:
  Applied Physics 36~(15) (2003) 1798--1805.
\newblock \href {https://doi.org/10.1088/0022-3727/36/15/309}
  {\path{doi:10.1088/0022-3727/36/15/309}}.

\bibitem{alics}
S.~Ali, Electron-ion recombination data for plasma applications: Results from
  electron beam ion trap and ion storage ring, Ph.D. thesis, Stockholm
  University, Department of Physics (2012).

\bibitem{birvhs}
G.~A. Bird, {Monte Carlo simulation in an engineering context}, {Progress in
  Astronautics and Aeronautics} 74 (1981) 239--255.
\newblock \href {https://doi.org/10.2514/5.9781600865480.0239.0255}
  {\path{doi:10.2514/5.9781600865480.0239.0255}}.

\bibitem{lapdsw}
G.~Lapenta,
  \href{https://www.sciencedirect.com/science/article/pii/S0021999102971263}{{Particle
  Rezoning for Multidimensional Kinetic Particle-In-Cell Simulations}},
  {Journal of Computational Physics} 181~(1) (2002) 317--337.
\newblock \href {https://doi.org/10.1006/jcph.2002.7126}
  {\path{doi:10.1006/jcph.2002.7126}}.
\newline\urlprefix\url{https://www.sciencedirect.com/science/article/pii/S0021999102971263}

\bibitem{teudsw}
J.~Teunissen, U.~Ebert,
  \href{https://www.sciencedirect.com/science/article/pii/S0021999113008048}{{Controlling
  the weights of simulation particles: adaptive particle management using k-d
  trees}}, {Journal of Computational Physics} 259 (2014) 318--330.
\newblock \href {https://doi.org/10.1016/j.jcp.2013.12.005}
  {\path{doi:10.1016/j.jcp.2013.12.005}}.
\newline\urlprefix\url{https://www.sciencedirect.com/science/article/pii/S0021999113008048}

\bibitem{mardsw}
R.~S. Martin, J.-L. Cambier,
  \href{https://www.sciencedirect.com/science/article/pii/S0021999116000280}{{Octree
  particle management for DSMC and PIC simulations}}, {Journal of Computational
  Physics} 327 (2016) 943--966.
\newblock \href {https://doi.org/10.1016/j.jcp.2016.01.020}
  {\path{doi:10.1016/j.jcp.2016.01.020}}.
\newline\urlprefix\url{https://www.sciencedirect.com/science/article/pii/S0021999116000280}

\bibitem{gondsw}
A.~Gonoskov,
  \href{https://www.sciencedirect.com/science/article/pii/S001046552100312X}{{Agnostic
  conservative down-sampling for optimizing statistical representations and PIC
  simulations}}, {Computer Physics Communications} 271 (2022) 108200.
\newblock \href {https://doi.org/10.1016/j.cpc.2021.108200}
  {\path{doi:10.1016/j.cpc.2021.108200}}.
\newline\urlprefix\url{https://www.sciencedirect.com/science/article/pii/S001046552100312X}

\bibitem{yamspuy}
Y.~Yamamura, H.~Tawara, Energy dependence of ion-induced sputtering yields from
  monatomic solids at normal incidence, Atomic Data and Nuclear Data Tables
  62~(2) (1996) 149--253.
\newblock \href {https://doi.org/10.1006/adnd.1996.0005}
  {\path{doi:10.1006/adnd.1996.0005}}.

\bibitem{stuspue}
R.~V. Stuart, G.~K. Wehner, {Energy Distribution of Sputtered Cu Atoms},
  Journal of Applied Physics 35~(6) (1964) 1819--1824.
\newblock \href {https://doi.org/10.1063/1.1713748}
  {\path{doi:10.1063/1.1713748}}.

\bibitem{caldfi}
S.~Calatroni,
  \href{https://indico.cern.ch/event/774138/contributions/3566528/}{{Direct
  Field Ionization}}, in: {8th International Workshop on Mechanisms of Vacuum
  Arcs}, 2019.
\newline\urlprefix\url{https://indico.cern.ch/event/774138/contributions/3566528/}

\bibitem{brudfi}
D.~L. Bruhwiler, D.~A. Dimitrov, J.~R. Cary, E.~Esarey, W.~Leemans, R.~E.
  Giacone, Particle-in-cell simulations of tunneling ionization effects in
  plasma-based accelerators, Physics of Plasmas 10~(5) (2003) 2022--2030.
\newblock \href {https://doi.org/10.1063/1.1566027}
  {\path{doi:10.1063/1.1566027}}.

\bibitem{kyrhea}
A.~Kyritsakis, M.~Veske, K.~Eimre, V.~Zadin, F.~Djurabekova, Thermal runaway of
  metal nano-tips during intense electron emission, Journal of Physics D:
  Applied Physics 51~(22) (May 2018).
\newblock \href {https://doi.org/10.1088/1361-6463/aac03b}
  {\path{doi:10.1088/1361-6463/aac03b}}.

\bibitem{ovito}
A.~Stukowski, {Visualization and analysis of atomistic simulation data with
  OVITO - the Open Visualization Tool}, {Modelling and Simulation in Materials
  Science and Engineering} 18~(1) (Jan. 2010).
\newblock \href {https://doi.org/10.1088/0965-0393/18/1/015012}
  {\path{doi:10.1088/0965-0393/18/1/015012}}.

\bibitem{rouexp}
A.~G. Rousskikh, A.~S. Zhigalin, V.~I. Oreshkin, R.~B. Baksht, Measuring the
  expansion velocity of plasma formed during electrical breakdown along an
  exploding al foil in a medium of desorbed gases, Journal of Physics:
  Conference Series 1393~(1) (2019) 012020.
\newblock \href {https://doi.org/10.1088/1742-6596/1393/1/012020}
  {\path{doi:10.1088/1742-6596/1393/1/012020}}.

\end{thebibliography}

\end{document}